\documentclass[10pt,journal,compsoc]{IEEEtran}
\usepackage{enumerate}
\usepackage{booktabs}
\usepackage{multirow}
\usepackage{algorithm}
\usepackage{algpseudocode}
\usepackage{comment}
\usepackage{amsmath,amssymb}
\usepackage{latexsym}
\usepackage{graphicx}
\usepackage{subfigure}
\usepackage{epstopdf}
\usepackage{esvect}
\usepackage{amssymb}
\usepackage{caption}
\usepackage{epsfig}
\usepackage{eqnarray,amsmath}
\usepackage{mathtools}
\usepackage{xspace}
\usepackage{xcolor}
\usepackage[bottom]{footmisc}
\usepackage[11pt]{moresize}
\usepackage{url}

\usepackage{enumitem}

\newcommand{\model}{\textbf{ODHD}}
\newcommand{\immodel}{\textbf{IM-ODHD}}
\newcommand{\mypara}[1]{
	\vspace*{0.01cm}
	\noindent\textbf{\textit{#1}}}
	
\newcommand{\rev}[1]{\textcolor{black}{#1}}
\newcommand{\review}[1]{\textcolor{black}{#1}}

\begin{document}	
\title{A Computing-in-Memory-based One-Class Hyperdimensional Computing Model \\for Outlier Detection}

% \author{%
%   \IEEEauthorblockN{%
%     Ruixuan~Wang\IEEEauthorrefmark{1}\textsuperscript{\textsection},
%     Sabrina~Hassan\IEEEauthorrefmark{2}\textsuperscript{\textsection},
%     X.~Sharon~Hu\IEEEauthorrefmark{3}
%     Xun~Jiao\IEEEauthorrefmark{1} and
%     Dayane~Reis\IEEEauthorrefmark{2}%
%   }%
%   \IEEEauthorblockA{\IEEEauthorrefmark{1} Villanova University}%
%   \IEEEauthorblockA{\IEEEauthorrefmark{2} Affiliation 2}%
%   \IEEEauthorblockA{\IEEEauthorrefmark{3} Affiliation 3}%
% }

% \maketitle
% \begingroup\renewcommand\thefootnote{\textsection}
% \footnotetext{Equal contribution}
% \endgroup

\author{Ruixuan~Wang\textsuperscript{\textsection},\IEEEmembership{ Student Member,~IEEE},
        Sabrina~Hassan~Moon\textsuperscript{\textsection},\IEEEmembership{ Student Member,~IEEE},
        X.~Sharon~Hu,~\IEEEmembership{Fellow,~IEEE,}
        Xun~Jiao,~\IEEEmembership{Member,~IEEE,}
        and Dayane~Reis,~\IEEEmembership{Senior Member,~IEEE}

\IEEEcompsocitemizethanks{   

\IEEEcompsocthanksitem Corresponding authors: Xun Jiao (xun.jiao@villanova.edu) and Dayane Reis (dayane3@usf.edu).
\IEEEcompsocthanksitem R. Wang and X. Jiao are with the Department of Electrical and Computer Engineering, Villanova University.\protect
\IEEEcompsocthanksitem S. H. Moon and D. Reis are with the Department of Computer Science and Engineering, University of South Florida.\protect
\IEEEcompsocthanksitem X. S. Hu is with the Department of Computer Science and Engineering, University of Notre Dame.% <-this % stops a space

\textsection{ The authors contributed equally to this work.}

}%end thanks block
}%end author block      

\IEEEtitleabstractindextext{%
\begin{abstract}
In this work, we present \model, an algorithm for outlier detection based on hyperdimensional computing (HDC), a non-classical learning paradigm. Along with the HDC-based algorithm, we propose \immodel, a computing-in-memory (CiM) implementation based on hardware/software (HW/SW) codesign for improved latency and energy efficiency. The training and testing phases of \model~may be performed with conventional CPU/GPU hardware or our \immodel, SRAM-based CiM architecture using the proposed HW/SW codesign techniques. We evaluate the performance of \model ~on six datasets from different application domains using three metrics, namely accuracy, F1 score, and ROC-AUC, and compare it with multiple baseline methods such as OCSVM, isolation forest, and autoencoder. The experimental results indicate that \model ~outperforms all the baseline methods in terms of these three metrics on every dataset for both CPU/GPU and CiM implementations. Furthermore, we perform an extensive design space exploration to demonstrate the tradeoff between delay, energy efficiency, and performance of \model. We demonstrate that the HW/SW codesign implementation of the outlier detection on \immodel~is able to outperform the GPU-based implementation of \model~by at least \review{331.5$\times$/889}$\times$ in terms of training/testing latency (and on average \review{14.0$\times$/36.9$\times$} in terms of training/testing energy consumption). 
\end{abstract}

}

\maketitle

\pagestyle{plain}

\section{Introduction}
Outlier detection, also referred to as anomaly detection, is a crucial technique utilized in various application domains like medical diagnosis, Internet-of-Things (IoT), and financial fraud detection. Outliers are generally extreme or out-of-distribution values in a dataset that deviate from other samples or an observation that does not fit the overall pattern. These outliers typically suggest measurement variability, experimental errors, or novelty. In machine learning, outliers in the training or testing set may cause failure in the detection or classification tasks. Additionally, in recent times, cyber attackers deliberately fabricate outliers, posing a threat to the security of cyber-physical systems.

Over the years, researchers continue to design robust solutions to detect outliers efficiently and effectively, where statistical methods and machine learning methods are the two most popular types of solutions. Statistical methods include parametric methods such as Gaussian mixture model (GMM) methods~\cite{tang2015outlier} and non-parametric methods such as kernel density estimation methods~\cite{latecki2007outlier}. While statistical methods are mathematically well explainable, fast to evaluate, and easy to implement, their results could be unreliable for practical applications due to their dependency on assumptions of a specific distribution model.

Recently, using machine learning techniques for outlier detection has witnessed a significant surge. Among the most effective methodologies, one-class support vector machine (OCSVM), isolation forest, and autoencoder-based neural network approaches are the most notable. OCSVM, which is a variant of the conventional SVM, distinguishes outliers from inliers by maximizing the margin~\cite{li2016anomaly}. The isolation forest method, on the other hand, utilizes an ensemble model consisting of isolation trees, with outliers being more vulnerable to isolation and having shorter traversal path lengths~\cite{liu2008isolation}. The autoencoder-based method is a novel unsupervised learning approach that uses neural networks to reconstruct data samples, identifying outliers based on the reconstruction errors~\cite{he2020exploring}. These traditional machine learning-based methods can achieve accurate outlier detection but may lack consideration for computation and energy efficiency.

In this paper, we present a novel approach for outlier detection based on hyperdimensional computing (HDC). HDC is an emerging computing paradigm inspired by the human brain circuitry that exhibits high-dimensionality and fully distributed holographic representation~\cite{kanerva2009hyperdimensional, ge2020classification}. HDC represents data samples using high-dimensional hypervectors, typically dimension $D=10,000$, which can be generated, manipulated, and compared to perform learning tasks. Compared to deep neural networks (DNNs), HDC offers several advantages, including smaller model size, lower computational cost, and one/few-shot learning, making it an attractive alternative, particularly for low-cost computing platforms~\cite{ge2020classification}. HDC has demonstrated promising results in diverse applications such as computer vision~\cite{hersche2020integrating}.
%and natural language processing~\cite{rahimi2016robust}. 

Specifically, we propose~\model, which is a novel one-class HDC-based outlier detection method using a positive-unlabeled (P-U) learning structure \cite{elkan2008learning}. Our approach is based on the simple yet reasonable assumption that a single hypervector (HV) can represent the abstract information of all inlier samples, which can be distinguished from outlier samples represented in HVs. Although HDC has been extensively studied for supervised learning tasks such as classification in various domains~\cite{ge2020classification}, there is limited research on using HDC for other tasks. Furthermore, recognizing the memory-centric computing properties of \model, we propose a hardware/software codesign implementation of \model`s~both training and inference phases on top of a computing-in-memory (CiM) architecture (\immodel). CiM can curtail the memory access bottleneck by leveraging parallelism inside the memory array structure, which enables computation at the bitline level along several current paths simultaneously. CiM has emerged as one of the most promising approaches for signal processing, optimization, deep learning and stochastic computing~\cite{ielmini2020device}. Our experimental results indicate that CiM can significantly accelerate the \model~algorithm and deliver superior energy efficiency.
% Furthermore, we implement \model ~on top of a computing-in-memory (CiM) architecture, which conforms to the unique memory-centric computing properties of \model. CiM can curtail the bottleneck by leveraging parallelism inside the memory array structure, which enables computation at the bitline level along several current paths simultaneously. CiM has emerged as one of the most promising approaches for medical applications~\cite{penkovsky2020memory}, big data analytics acceleration in health care~\cite{patel2014big}, signal processing, optimization, deep learning and stochastic computing~\cite{ielmini2020device}. Here, we introduce a CiM architecture that leverages hardware/software codesign techniques to implement the training and inference phases of \model. Results show that CiM can significantly accelerate the \model ~process and promote superior energy efficiency. \textbf{Specifically, built on top of our previous study in~\cite{wang2022odhd}, this paper makes the following contributions: }

Built on top of our previous study in~\cite{wang2022odhd}, this paper makes the following contributions:

\begin{enumerate}
    \item We introduce \model, a novel one-class outlier detection method based on HDC and P-U learning. Our approach forms a high-dimensional representation of inlier samples and is a viable alternative to existing outlier detection approaches. 
    \item We develop a comprehensive pipeline for \model~algorithm. First, we map all inliers samples to a high-dimensional space and create a one-class HV to represent the abstract information of inliers. Next, we propose a confidence-based method to automatically compute a threshold that is used for outlier detection. During testing, we compute the similarity between the unseen testing sample and the one-class HV and compare it to the pre-computed threshold to detect outliers.
    \item We propose a static random-access memory (SRAM)-based CiM architecture, \immodel, to implement \model. Our CiM architecture leverages customized elements (sense amplifiers), mat-level row/column decoders, logarithmic bit shifters, etc., to attain reduced latency and increased parallelism supporting different parameters of HDC, such as a number of dimensions and hypervector seeds. 
    \item We apply a hardware/software codesign approach to further improve the functionality of \immodel~on CiM architecture. Specifically, we adjust the algorithm-level design of \model ~and show that the proposed changes speed up the runtime of both training and inference with insignificant accuracy loss. 
\end{enumerate}

To evaluate \model, we use six datasets from the Outlier Detection Datasets (ODDS) Library~\cite{Rayana2016} and compare our approach with baseline methods such as OCSVM, isolation forest, autoencoder, and HDAD. The comprehensive evaluation results show that \model ~outperforms all the baseline methods on all six datasets in all metrics, including accuracy, F1 score, and ROC-AUC with both CPU/GPU and CiM implementations. Furthermore, after the hardware/software codesign adjustment, we demonstrate that our hardware/software codesign implementation of \immodel~is able to outperform the GPU-based implementation of the same algorithm by at least 293$\times$/\rev{419}$\times$ in terms of training/testing latency (and on average \rev{16.0$\times$/15.9$\times$} in terms of training/testing energy consumption). Our study demonstrates the effectiveness of \model~in the realm of both software and hardware and highlights its potential for research in outlier detection.

The rest of the paper is structured as follows. In Sec.~\ref{sec:background}, we discuss the fundamentals of HDC and CiM. Sec.~\ref{sec:algo} introduces the \model~algorithm. An SRAM-based CiM architecture for \model~(\immodel) is proposed in Sec.~\ref{sec:hw}. We evaluate the performance, energy, and latency of \model~ and \immodel~in Sec.~\ref{sec:exp}. Related works are presented and discussed in Sec.~\ref{sec:related_work}. Finally, Sec.~\ref{sec:conclusion} concludes the paper.
\section{Background}

\label{sec:background}
Here, we discuss the mathematical foundations and operations of HDC. Furthermore, we discuss the basics of CiM.
\subsection{Hyperdimensional Computing}

\mypara{Basic HDC Component:}
Hypervectors (HVs) are the fundamental components of HDC. An HV is a holographic and high-dimensional vector with independent and identically distributed (i.i.d.) elements. The HV with $D = d$ dimensions is denoted as $\vv{H} = \langle h_1, h_2, \dots, h_d\rangle$. In this paper, we employ bipolar HVs, which means each element in an HV is either~$-1$~or~$1$~\cite{ge2020classification}. 

In HDC, HVs are used as the information representation in different scales and levels, such as embedding new information or aggregating existing information. To measure the correlation between information representation, we use cosine distance to measure the similarity of information between two HVs, as shown in Eq.~\ref{eqn:sim}. Moreover, one property of HVs is, when the dimensionality is sufficiently high (e.g., $D = 10,000$), HVs are quasi-orthogonal whereas any two random bipolar HVs are nearly orthogonal~\cite{kanerva2009hyperdimensional}. 
\begin{equation}
\small
    \delta(\vv{H_x}, \vv{H_y}) = \frac{\vv{H_x} \cdot \vv{H_y}}{||\vv{H_x}||\times||\vv{H_y}||} = \frac{\sum_{i=1}^{d} h_{xi} \cdot h_{yi}}{\sqrt{\sum_{i=1}^{d} {h_{xi}}^2} \cdot \sqrt{\sum_{i=1}^{d} {h_{yi}}^2}}
\label{eqn:sim}
\end{equation}

\mypara{Basic HDC Operations:} 
HDC supports three basic arithmetic operations including bundling, binding and permutation, as illustrated in Eq.~\ref{eqn:operation}. Additions and multiplications both take two input HVs as operands and perform \textbf{element-wise} add or multiply operations. Permutation takes one HV as the input operand and performs \textbf{cyclic rotation}. 
\begin{equation}
\begin{aligned}
bundling(\vv{H_x} + \vv{H_y}) & = \langle h_{x1} + h_{y1}, h_{x2} + h_{y2}, \dots, h_{xd} + h_{yd}\rangle \\
binding(\vv{H_x} * \vv{H_y}) & = \langle h_{x1} * h_{y1}, h_{x2} * h_{y2}, \dots, h_{xd} * h_{yd}\rangle \\
permutation^1(\vv{H}) & = \langle h_d, h_1, h_2, \dots, h_{d-1}\rangle 
\end{aligned}
\label{eqn:operation}
\end{equation}
All three operations preserve the dimensionality of the input HVs, i.e., the input HVs and the output HVs have the same dimension. Considering the three main operations, bundling adds the same type of information, binding aggregates various types of information together to generate new information, and permutation reflects the spatial or temporal changes, such as time series or spatial coordinates~\cite{kanerva2009hyperdimensional}.  

\subsection{Computing-in-Memory}
The limited processor-memory bandwidth significantly impacts a system's performance. Computing-in-memory (CiM) performs the logic and memory operations associated with a given task within the memory boundaries. CiM exploits the large, internal bandwidth of memory to achieve parallelism, which reduces latency and saves energy due to fewer external memory references. CiM architectures may target either general-purpose or application-specific designs, as described below. 

\subsubsection{Application-Specific CiM Designs}
Examples of application-specific CiM designs include the in-memory computation of dot-products with crossbars \cite{kang14} and search with non-volatile ternary content addressable memories (TCAMs) \cite{yin2018ultra}, which are suitable for performing nearest neighbor operations. The majority of CiM implementations for HDC rely on application-specific designs based on crossbars and TCAMs. These CiM architectures typically employ emerging memory technologies (EMTs) such as Ferroelectric Field-Effect Transistors (FeFETs) (e.g., \cite{kazemi2022achieving}) and Resistive Random-Access Memories (ReRAMs) (e.g., \cite{liu19_hdc_rram}). EMTs have great potential for high-density and low-power implementations of CiM-based HDC. For instance, CiM improves energy consumption by 826$\times$ and latency by 30$\times$ for a classification task with HDC when compared to a GPU baseline \cite{kazemi2022achieving}. However, as development on EMTs is still in its early phases, there is a lack of large-scale solutions for CiM-based HDC that can be promptly integrated into real systems. Furthermore, much of the computation with EMTs in application-specific CiM designs occurs in the analog domain, which limits the bit precision due to the physical limits of the EMTs, as well as the errors induced by circuit components such as the analog-to-digital converters (ADCs). The limited precision makes it challenging to match software accuracies.

\subsubsection{General-Purpose CiM Designs}
\label{sec:background_CiMcircuits}
General purpose CiM (GPCiM) designs support logic and arithmetic operations that can benefit different applications as they can be used to implement different algorithms~\cite{reis2018computing}. In this work, we propose a GPCiM architecture that is capable of performing all the operations needed by HDC-based outlier detection. The algorithmic flow for outlier detection with HDC running on our CiM architecture is presented in Sec. \ref{sec:algo}. The CiM architecture of \model, which we name as \immodel, is described in Sec. \ref{sec:hw}, along with a hardware/software codesign approach for \model~that allows for mapping of the outlier detection algorithm onto \immodel. \immodel~operates in the digital domain, with customized sense amplifiers, local copy drivers, and bit shifters, achieving high parallelism with multiple subarrays operating simultaneously to perform in-memory operations. The circuits employed in \immodel~ are illustrated in part~\textit{b} of Fig. \ref{fig:architecture} and described below.
%Examples of GPCiM designs include \cite{reis2018computing,li16}.

~\mypara{Word Line Decoders \cite{reis20_date}: }The simultaneous sensing of multiple rows in an SRAM subarray is possible by lowering the word line voltage to bias against the write of the SRAM. As shown in Fig.~\ref{fig:architecture}, to leverage double sensing, our design implements two-word line decoders in the same PE to simultaneously activate two rows for performing computation between them.

~\mypara{Customized Sense Amplifier (CSA) \cite{aga17}: }Once the subarray rows are activated, voltage drops on the memory bitlines (and negated bitlines), while the actual values depend on the operands stored in the SRAM. The voltage drop can be sensed with CSA, which will generate the results for different bitwise logic operations (e.g., AND, OR, XOR) and arithmetic between the two rows of data. The output of the CSA depends on the desired operation, which is selected with an internal multiplexer circuit. 

~\mypara{Logarithmic Bit Shifters \cite{reis20_date}: }The output of the CSA is passed as input to the logarithmic bit shifter, which can shift the output to the left or right. The number of bit positions by which a binary number is shifted left or right is determined by the logarithm of a shift amount (i.e., the shift mask in our circuit). The main advantage of this circuit over a traditional linear bit shifter is that it can perform larger shifts in a single clock cycle, rather than shifting one-bit position at a time. The logarithmic bit shifter in \immodel~can accelerate permutations and divisions by powers-of-two in \model. In the logarithmic bit shifter used in \immodel~(shown in Fig.~\ref{fig:architecture}), a 3-bit shift mask goes into each PE to configure 0-3 bit shifts to the left or right (1 bit of the mask determines the shift direction, while the other two bits are for the shift amount). \rev{Shift amounts larger than 3 bits are possible through a multi-step approach.}
%This implementation is shown in Fig. \ref{fig:architecture}(b). 

~\mypara{Write and Copy Drivers \cite{reis20_tvlsi}: }Memory write drivers are circuits that play a crucial role in writing data into the 6T-SRAM memory cells. These drivers work by amplifying the signals from an external memory controller to generate the required voltage levels for writing data into the memory cells through the bit lines. In the context of the \model, write drivers are utilized to write both the initial HVs and intermediated HVs from the mat-level registers into the PEs. Copy drivers are another type of circuitry used in \model~specifically for copying the results of CiM operations, such as bitwise logic, addition, or right/left bit shift, to a designated address within the same PE. To ensure efficient and effective copying, copy drivers are placed in alignment with the CSA columns in the subarray.

%Furthermore, as computation occurs \textit{in loco} without data stored in a centralized memory unit such as the dynamic random access memory (DRAM), it is possible to eliminate energy and latency overheads of external memory references with our architecture. \model~requires a hardware/software codesign approach for the outlier detection algorithm also described in Sec. \ref{sec:hw}, which ensures high accuracy while fully exploiting the capabilities of our CiM design for the short delay and high energy efficiency.

%Computing-in-Memory (CiM) is a computer architecture paradigm technique of performing computing actions within the architecture of memory systems for reducing execution time and increasing operational efficiency. It enables faster computation and reduced I/O latency and power consumption, as opposed to conventional disk storage, by using the RAM of a computer to store and process data. The main factor of bottleneck in data processing is the time of data transmission between memory and the processor which is minimized in in-memory computing \cite{8811809}. Due to how quickly data can be read from and written to memory than to a disk, this enables substantially faster data access and processing speeds. This approach is particularly useful for big data and machine learning applications, where large amounts of data need to be processed quickly and efficiently.

\begin{figure*}[t]
    \centering
    \includegraphics[width = 1.5\columnwidth]{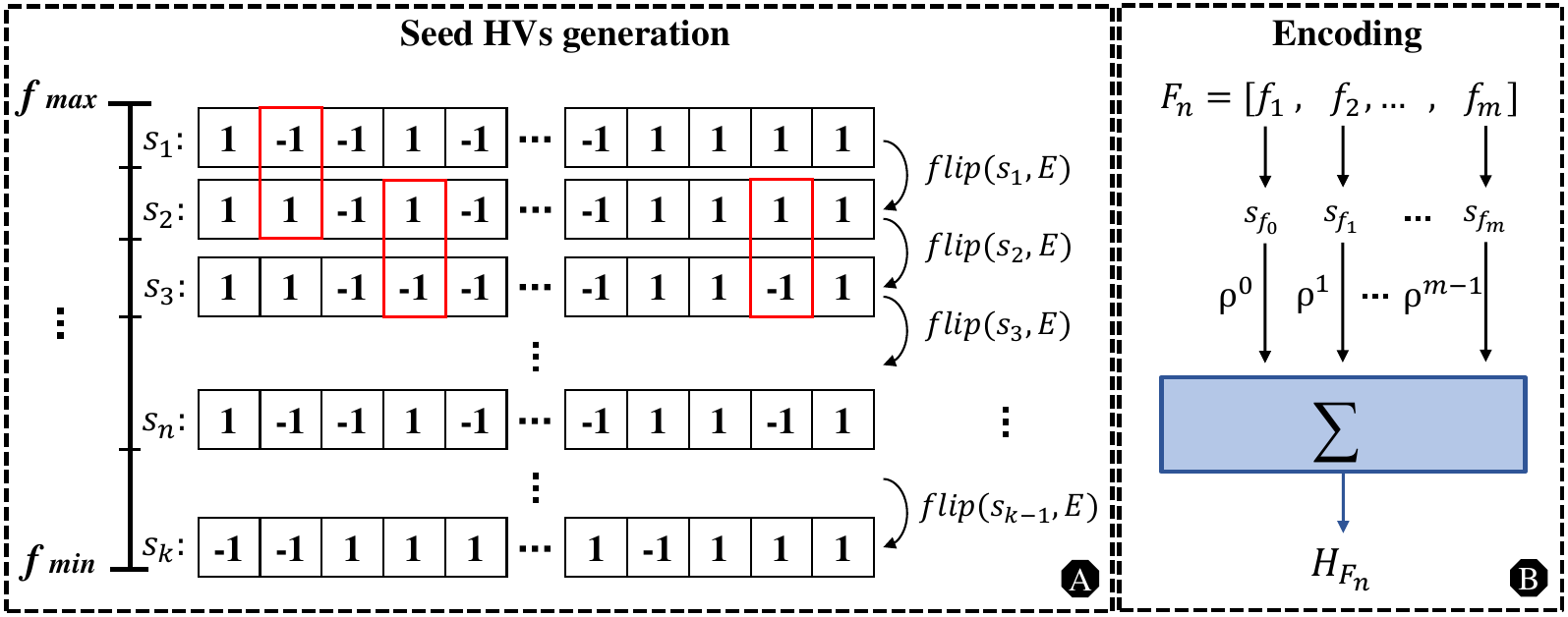}
    \includegraphics[width = 1.5\columnwidth]{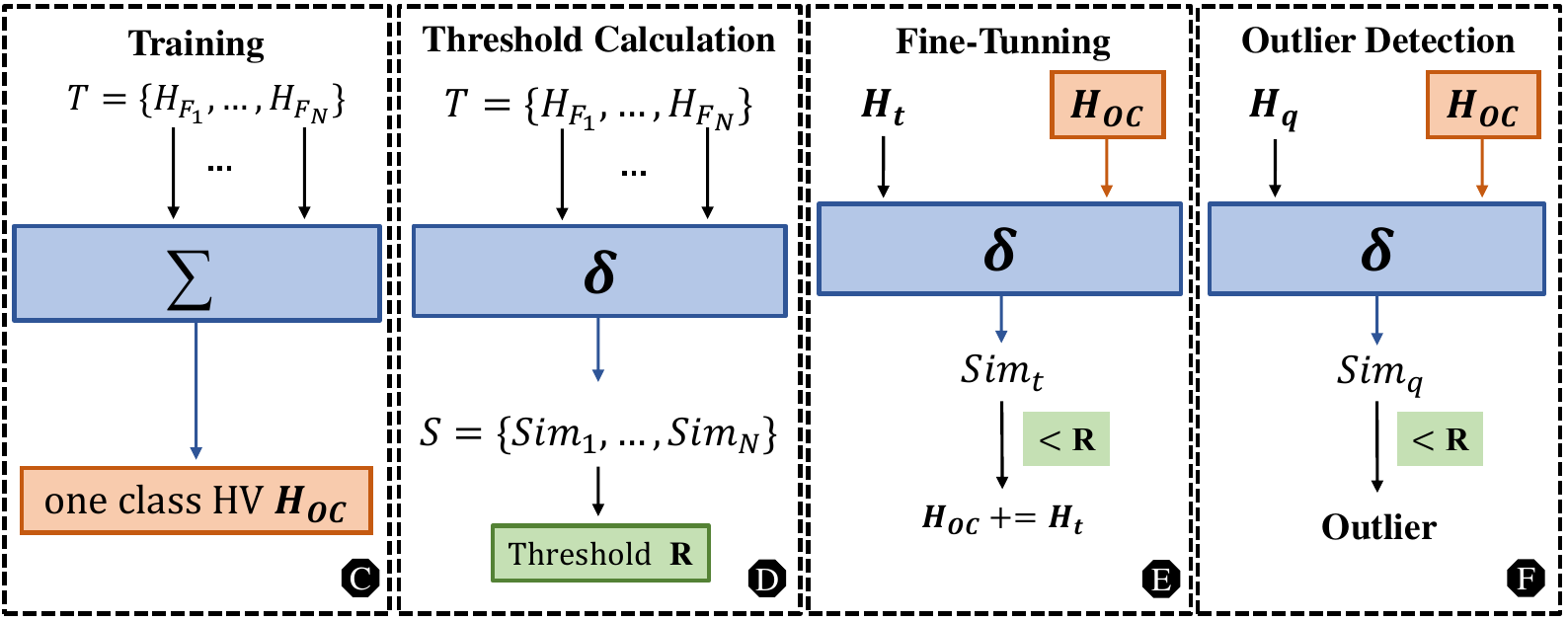}

	\caption{The algorithmic flow of \model~ with six key phases.}
    \label{fig:framework}

\end{figure*}

\section{\model: Algorithm}
\label{sec:algo}
%In this section, we describe the algorithmic flow of \model. 
We leverage the mathematical properties of HDC to develop a novel one-class HDC-based outlier detection algorithm, which essentially learns an abstract representation of inlier samples and then performs one-class classification-based outlier detection. In \model, the outlier detection process is based on a P-U learning structure~\cite{elkan2008learning}, which means we use only inlier samples for training and test on a testing set (may contain both inliers and outliers) without the information of labels. The one-class HV we trained contains the information from all the patterns of inlier (training) samples. For inference, we detect whether a query HV conforms to the one-class HV according to cosine similarity. In \model, we utilize a confidence-based procedure to calculate a threshold based on training HVs. If the cosine similarity between a query HV and one-class HV is lower than the threshold, the query HV will be detected as an outlier. Fig.~\ref{fig:framework} illustrates the whole algorithmic flow of \model, which is divided into six key phases: Seed HV Generation, Encoding, Training, Threshold Calculation, Fine-Tuning, and Outlier Detection. We describe each phase of the algorithm in detail in the following sections.

\subsection{Seed HVs Generation}
\label{sec:seed}
As the first step, we need to generate seed HVs so that we can encode the raw sample features into HVs. 
As noted previously, each HV is a high-dimensional vector with i.i.d elements~\cite{kanerva2009hyperdimensional}. We employ an HV-generating method consistent with the one in~\cite{kim2018efficient} to create $k$ seed HVs that can support later encoding, which is more computationally efficient compared to randomly generating $k$ random HVs straightforwardly while preserving the orthogonality of HVs. As part A of Fig.~\ref{fig:framework} illustrates, we initiate a random bipolar $D$-dimension HV, $\vv{s_1}$, and then generate all the seed HVs by randomly flipping $E = D/2k$ elements. Specifically, $k$ is a configurable parameter depending on how we discretize the input data. Consequently, a set of seed HVs $\{\vv{s_1}, \vv{s_2}, \dots, \vv{s_k}\}$ is generated. For the following encoding procedure, assume for a specific dataset, we have each feature vector with $m$ feature elements $\vv{F_n} = \langle f_1, f_2, \dots, f_m \rangle$. According to the training set, we can capture the minimum and the maximum values of each feature value $f_{min}$ and $f_{max}$. Then we can discretize the input feature space $(f_{min}, f_{max})$ into $k$ uniform intervals. Thus, each feature value corresponds to a specific interval, and we can map the feature vector into an integer vector for encoding.
% A larger $k$ means that we have more quantization levels, and hence we can more precisely represent the original data but also increase the memory overhead (will discuss in Section.\ref{sec:space}). 

\subsection{Encoding}
\label{sec:encode}
%In this section, we explain the progress of encoding feature vectors into HVs by using the seed HVs we generated. 
The encoding step projects the original feature vector into an HV. 
The encoding process of feature vector $\vv{F_n} = {\langle f_1, f_2, \dots, f_m\rangle}$ is shown in part B of Fig.~\ref{fig:framework}. We first index the seed HV corresponding to each feature value. For example, if the feature element $f_2$ falls into the $5^{th}$ interval among the $k$ intervals, the corresponding seed HV is the $5^{th}$ of the $k$ seed HVs. Then, we employ the permutation operation to embed the information of the feature position into the seed HV. As the permutation operation reflects the spatial change of information, we bundle the information of feature position by deploying a cyclic rotation on each seed HV as shown in Eq.~\ref{eqn:operation}. Particularly, we keep the first seed HV un-permuted ($\rho^0(\vv{s_{f_1}})$), and for seed HV $\vv{s_2}$ to $\vv{s_k}$, we circularly rotate the $i^{th}$ seed HV by $i - 1$ elements, i.e., $\rho^{i-1}(\vv{s_{f_i}})$.

At the end of the encoding process, we aggregate all permuted seed HVs corresponding to all feature values into one HV $\vv{H_{F_n}}$ representing the entire feature vector $\vv{F_n}$. Note that if we have 100 inlier samples (i.e., 100 feature vectors) in the training dataset, we would have 100 corresponding encoded HVs. The overall encoding process is denoted as Eq.~\ref{eqn:encode}. 
\begin{equation}
    \begin{aligned}
    \vv{H_{F_n}} = \rho^0(\vv{s_{f_1}}) + \rho^1(\vv{s_{f_2}}) + \dots + \rho^{m-1}(\vv{s_{f_m}})
\end{aligned}
\label{eqn:encode}
\end{equation}
\subsection{Training}
\label{sec:train}
\begin{figure*}[t]
     \centering

     {
         \includegraphics[width=2\columnwidth]{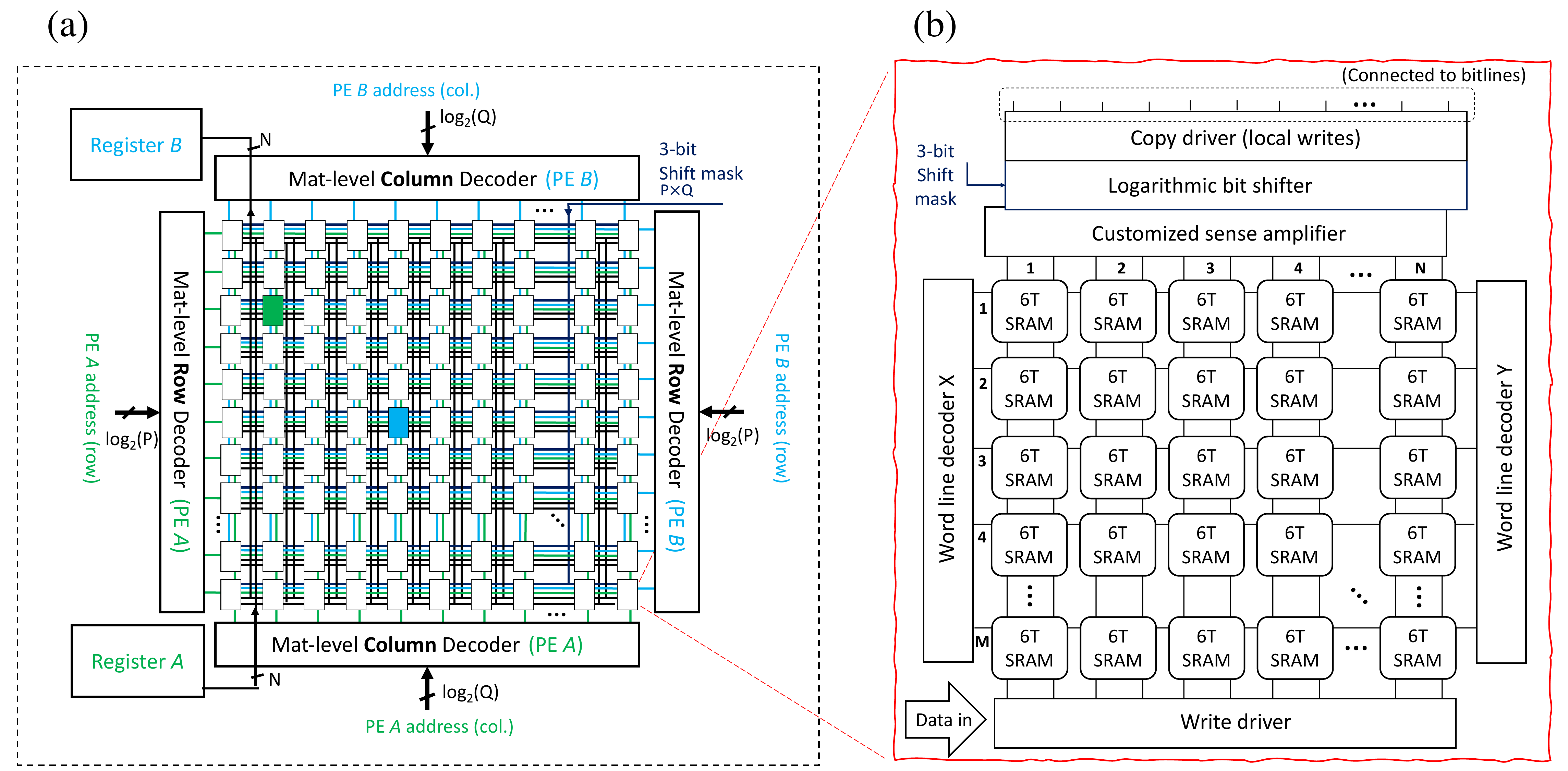}
     }
     %\hspace{-5pt}

 	\caption{IMC architecture for outlier detection. (a) ($P \times Q$) mat-level architecture. The PEs are marked green for source and blue for destination. (b) Detail of one ($M \times N$) subarray. }
 	\label{fig:architecture}
 \end{figure*}

As part C in Fig.~\ref{fig:architecture} indicates, after encoding all feature vectors in the training set, the training phase generates the one-class HV ($H_{OC}$) of the entire training set, i.e., all inlier samples. 
%, we implement the training process and generate the one-class HV $\vv{HV_{OC}}$. 
%Note that as the training set only contains inlier samples, \model ~is essentially a semi-supervised learning method. 
%Hence, as part C of Fig.~\ref{fig:framework} shows, we can use one HV to represent the whole training set by aggregating all training HVs directly. 
Eq.~\ref{eqn:train} illustrates the process of HDC training, which bundles all the HV representing each inlier feature vector. For example, if there are 100 inlier samples, then the 100 corresponding encoded HVs generated by the encoding process are added together to generate a single one-class HV $H_{OC}$ representing inlier samples or patterns.
\begin{equation}
    \begin{aligned}
    \vv{H_{OC}} = \sum_{i = 1}^{N} \vv{H_{F_i}}
    \end{aligned}
\label{eqn:train}
\end{equation}
\subsection{Threshold Calculation}
\label{sec:threshold}
In ODHD, we propose a confidence-based threshold calculation approach. In order to calculate a threshold to separate inliers and outliers, we measure the cosine similarity between $\vv{H_{OC}}$ and all training HVs to obtain a similarity array $S$. As part D in Fig.~\ref{fig:framework} shows, each similarity $Sim_i$ in array $S$ can be considered as the confidence of the training HV to be an inlier sample.

We calculate the mean value $\mu(S)$ and the standard deviation $\sigma(S)$ \rev{over all the similarity values in array $S$.  We then deploy the threshold estimation strategy shown in Eq.~\ref{eqn:thres}, which is established in prior research \cite{he2020exploring, wang2021brief}.}
\begin{equation}
    R = \mu(S) + 2*\sigma(S)
    \label{eqn:thres} 
\end{equation}
Ultimately, we compute the threshold $R$ based on the confidence of all training HVs. In the outlier detection domain, only the samples with cosine similarity higher than the threshold are determined as an inlier, while all the samples with cosine similarity lower than the threshold are identified as outliers.

%Figure below is from the hardware section:

\subsection{Fine-Tuning}
\label{Sec:FT}

%During fine-tuning, we employ the threshold $R$ to validate the confidence of each training HV. 
After the training phase, we expect that all training HVs should be properly determined as inliers (but this may not be the case). \rev{Hence, we perform fine-tuning for ODHD to enhance the performance of outlier detection. The fine-tuning process, shown in part E of Fig.~\ref{fig:framework}, is automatically conducted via pre-defined rules consisting of two steps: (1) measure the similarity metric between the encoded training HVs and the one-class HV; (2) if this similarity metric falls below the threshold calculated in step (D) of Fig. \ref{fig:framework}, incorporate the training HV into the one-class HV.}

\rev{The fine-tuning process acts as an auto-calibration process, and the single parameter that needs to be set by the user is the number of epochs for the fine-tuning process, e.g., in this paper, we executed the fine-tuning process for a total of 10 epochs. Note that we still only use the given training dataset for the fine-tuning process.}
In each fine-tuning epoch, we feed all the training samples to ODHD. For each training sample, we estimate the cosine similarity $Sim_{t}$ between the training HV $\vv{H_t}$ and one-class HV $\vv{H_{OC}}$. If $Sim_{t}$ is higher than threshold $R$, which means the estimation is correct, we do not make any changes to $\vv{H_{OC}}$. 
However, if $Sim_{t}$ is lower than $R$, which means ODHD mistakenly considers the inlier sample $t$ as an outlier, we update the one-class HV: we add the misclassified training HV $\vv{H_t}$ into the one-class HV to update the corresponding information in $\vv{H_{OC}}$. 
% For consistency, we perform 10 fine-tuning epochs for all cases.
%At the end of the retraining process, we dynamically update the threshold for further retraining. 
% \begin{equation}
%     \begin{aligned}
%     \vv{H_{OC}} = \vv{H_{OC}} + \vv{H_t}
%     \end{aligned}
% \label{eqn:retrain}
% \end{equation}
\subsection{Outlier Detection}
\label{sec:detect}
After we train $\vv{H_{OC}}$ and obtain threshold $R$ based on the confidences of the training HVs, we deploy the outlier detection on an unseen sample without knowledge of the labels. The outlier detection process is shown as part F in Fig.~\ref{fig:framework}. 

During the outlier detection phase, we encode testing sample $q$ into an HV called query HV, $\vv{H_q}$, following the same encoding process in Eq.~\ref{eqn:encode} based on the same seed HVs. Then we compute the cosine similarity $Sim_{q}$ between the query HV $\vv{H_q}$ and the one-class HV $\vv{H_{OC}}$. In the event that $Sim_{q}$ is lower than the pre-computed threshold, the sample $q$ will be determined as an outlier.
\begin{equation}
\vv{H_q} = 
\begin{cases}
 Inlier& Sim_{q} \geq R\\
 Outlier& Sim_{q} < R
\end{cases}
\label{eqn:quantz}
\end{equation}

\section{\immodel: Hardware}
\label{sec:hw}

In this section, we first describe \immodel, the CiM-based hardware architecture for \model ~(Sec. \ref{sec:architecture}). We then discuss our hardware-algorithm codesign effort in adjusting \model~to the \immodel~hardware (Sec. \ref{sec:hwsw_codesign}).

\subsection{GPCiM Architecture}
\label{sec:architecture}

The combination of HDC and CiM can be particularly beneficial since HDC operations involve the manipulation of holographic HVs, which can be performed efficiently in memory. By performing HDC operations in memory using CiM, it is possible to achieve significant improvements in performance and energy efficiency compared to traditional von Neumann architectures (as demonstrated in \cite{kazemi2022achieving, liu19_hdc_rram}).

In this work, to implement the HDC-based outlier detection algorithms, we depart from the use of application-specific CiM designs based on NVM. Instead, we design \immodel~as a general-purpose CiM architecture based on CMOS (i.e., with 6T-SRAMs). The use of an SRAM-based design instead of an NVM-based one leads to several advantages: (1)~Our architecture can perform both the training and testing phases of HDC-based outlier detection in memory since SRAM has a much lower writing cost than NVMs. (2)~Easier prototyping and fabrication, as CMOS is readily available as opposed to NVMs. (3)~Computation in the digital domain which reduces the need for sophisticated peripherals such as ADCs, DACs, and current-based programming circuits. (4)~Our general-purpose CiM architecture has the ability to easily accommodate changes in the algorithm (as long as implementing them only requires the same key operations of HDC, which are binding, bundling, and permutation). \rev{(5) Prior research on system-level integration and compiler support for CiM architectures, such as \cite{fujiki2019duality}, could be readily leveraged to bolster the integration of our CiM architecture into a broader computing stack as it can support all the operations realized by \immodel.}

\immodel~is depicted in Fig.~\ref{fig:architecture}. The design contains $P \times Q$ subarrays, each of which acts as a processing element (PE) in the CiM architecture. Each PE contains $M \times N$ SRAM cells. The tile-styled architecture enables high throughput for the HDC operations (binding, bundling, permutation) due to parallel computation across the different subarrays. The elements of the mat-level design (depicted in Fig. \ref{fig:architecture}(a)) are explained in Sec. \ref{sec:hardware_mat}. The subarray design with its storage and computing capabilities is discussed in detail in Sec. \ref{sec:hardware_subarray}. 

\subsubsection{Mat Design}
\label{sec:hardware_mat}

Our CiM architecture implements \rev{the following new elements --- decoders, registers, and buses} --- which enable computation at the mat level. Below, we describe each component \rev{of the architecture} in detail.

~\mypara{Decoders: }\rev{Decoders orchestrate data access and facilitate data movement across the different PEs in our \immodel~fabric. For instance, an example of PE~\textit{A} and PE~\textit{B} is given in Fig. \ref{fig:architecture}(a) by the tiles colored green and blue, respectively). PE~\textit{A} and PE~\textit{B} can be accessed concurrently using two pairs of decoders. Two ($log_2P+log_2Q$)-bit addresses are used to activate each of the PEs~\textit{A} and~\textit{B}. To access the PE~\textit{A}, we divide its ($log_2P+log_2Q$)-bit address into two parts; the $log_2P$ most significant bits of the address are used as the input to the row decoder (\textbf{Mat-level Row Decoder (PE~\textit{A})} in Fig. \ref{fig:architecture}(a)), and the $log_2Q$ least significant bits of the address are used as the input to the column decoder (\textbf{Mat-level Column Decoder (PE~\textit{B})} in Fig. \ref{fig:architecture}(a)). Analogously, when accessing the PE~\textit{B}, the $log_2P$ bits of its address are used as the input to the \textbf{Mat-level Row Decoder (PE~\textit{B})} in Fig. \ref{fig:architecture}(a), while the $log_2Q$ bits of the address are used as the input to the \textbf{Mat-level Column Decoder (PE~\textit{B})} in Fig. \ref{fig:architecture}(a). }

~\mypara{Registers: }After the decoders select \rev{PE~\textit{A} and PE~\textit{B}}, the data from the output of each PE is transferred to its corresponding register, either \rev{register~\textit{A} or register~\textit{B}} (as illustrated in Fig. \ref{fig:architecture}(a)). The data traffic between each PE and the \rev{registers~\textit{A} and~\textit{B}} is managed via two dedicated buses, which will be elaborated upon in the subsequent paragraph. Once the data has been stored in either \rev{register ~\textit{A} or~\textit{B}}, it can be rerouted back to any PE through a reverse pathway, which is leveraged by the permutation operations are used in the encoding phase of \model. Sec.  \ref{sec:hwsw_codesign} provides details about performing this step with \immodel.

\rev{~\mypara{Buses A and B: }}Our CiM design employs dedicated \rev{buses~\textit{A} and~\textit{B}} to support (1) data movement \rev{from/to} the PEs \rev{to/from registers~\textit{A} and~\textit{B}}, and (2) the setup of a bit shift amount \rev{and direction} for each PE so reads, divisions, and multiplications by powers-of-two are possible with our \immodel~fabric. To achieve (1), our proposed CiM architecture has two separate sets of $N$-bit wide \rev{buses~\textit{A} and~\textit{B}} that connect each PE to the \rev{registers~\textit{A} and~\textit{B}}. The width of the buses is chosen as $N$ so it matches the dimensions of an individual $M \times N$ PE. Furthermore, two pairs of \rev{selector lines} come out of the row/column decoders and spread through the \rev{$P \times Q$} PEs on the mat (see green and blue wiring in Fig. \ref{fig:architecture}(a)). These lines are used to select the \rev{PE~\textit{A} and the PE~\textit{B}} for data transfer. Note that only \rev{two tiles, i.e., PE~\textit{A} and the PE~\textit{B}, can be selected at a given time through each pair of decoders}, which avoids data conflicts on \rev{buses~\textit{A} and~\textit{B}}. For~(2), our proposed CiM architecture implements a $3 \times P \times Q$-bit wide bus on which the bit shift amounts used at each PE can be set up individually (more details about bit shifts with our CiM architecture can be found in Sec. \ref{sec:hardware_subarray}). %through a 2-bit shift mask that allows for 0-3 bit shifts to the left or right with logarithmic bit shifts placed at the subarray level. Note that this 2$\times$P$\times$Q bus enables distinct bit shift amounts to be used with each array simultaneously.

\subsubsection{Subarray Design}
\label{sec:hardware_subarray}

Subarrays are the fundamental PEs of our design with their merged storage and processing capabilities. Fig. \ref{fig:architecture}(b) illustrates our SRAM-based processing element (PE), which utilizes 6T-SRAM memory cells, word line decoders \rev{X and Y}, customized sense amplifiers, write and copy drivers, and logarithmic bit shifters. These components, akin to \cite{aga17, reis20_date, reis20_tvlsi}, are crucial for facilitating the necessary bundling, binding, and permutation operations required by \model. For an in-depth understanding of the role each component plays within the PE, readers can refer to Sec. \ref{sec:background_CiMcircuits}.

\rev{Importantly, besides building on these established PE structures, our work introduces near-memory computing (NMC) circuits at the mat level, such as the buses and auxiliary registers managed by decoders (described in Sec.~\ref{sec:hardware_mat}). The introduced NMC circuits enhance our design's ability to carry out permutations --- a feature uniquely tailored to \model's encoding phase that represents a departure from previous in-SRAM computing solutions. The introduction of NMC elements in our CiM architecture sets our work apart, as existing in-SRAM architectures do not address the challenge of data movement between CiM PEs.}

 \subsection{Hardware/Software (HW/SW) Codesign}
 \label{sec:hwsw_codesign}

This section explains the efficient mapping of the steps of the \model~algorithm~(Sec. \ref{sec:algo}) to the CiM architecture of \immodel~(presented in Sec. \ref{sec:architecture}). The mapping process adopts the HW/SW codesign principle to adjust the \model~algorithm to better utilize the capabilities of \immodel. 
%As highlighted in \cite{odhd_wang_dac2022}, the algorithm of \model~ is superior to other existing models for detecting outliers in terms of accuracy, F1-score, and AUC. %However, as encoding the features consumes 72.3\% of the training time and 82.5\% of the testing time on average, \model~exhibits longer execution time than other baseline models. 
HW/SW codesign enforced in CiM architecture can significantly increase the performance of \model~while having high accuracy, F1-score, and AUC, as evaluated and discussed in Sec. \ref{sec:exp}.

\subsubsection{Seed HVs Generation in \immodel}
\label{sec:hw_sw_seedHV}

The initial step of creating the seed HVs involves generating them externally using random bit flips. However, once $k$ seed HVs, each with $D$ dimensions, are produced, they get distributed across the $P \times Q$ PEs of the \immodel~fabric. The PEs have a size of $M \times N$, where $M$ is the number of rows, and $N$ is the number of columns. When a $D$-dimensional seed HV is distributed across the PEs, its elements are indexed to the row $i$ of each PE, where $i \epsilon [1, M]$. The storage of all the elements within each seed HV spreads across $D/N$ PEs of the \immodel~fabric. Hence, to store $k$ seed HVs, the size of the GPCiM architecture needed is $k \times D$, i.e. $P \times Q \times M \times N \geq k \times D $. \rev{This $k \times D $ segment of the \immodel~fabric is designated for our seed HV storage and remains unaltered throughout the computation.}

\subsubsection{Encoding in \immodel}
\label{sec:hw_sw_encoding}

Once the $k$ seed HVs are written to the \immodel~fabric, the next step is to encode a given feature vector into an HV. The encoding step involves applying \textbf{permutation} and \textbf{bundling} operations. \textbf{Permutation on \immodel} is implemented with circular shifts. The process of performing a circular shift is \rev{implemented in two rounds,} as follows. 
 
 {\rev{\textbf{Round 1 (R1): }}}Assume a $D$-dimensional \rev{seed} HV is mapped to the $i^{th}$ row of $D/N$ PEs. We simultaneously access $i^{th}$ row of the $D/N$ PEs holding the HV, \rev{and refer to the PEs} as the \rev{\textit{destination} and \textit{source} PEs, in an alternate fashion}. The $i^{th}$ row data at the source PE undergoes a bitwise AND operation with a pre-stored mask filled with 1`s at the $m$ least significant bit positions and 0`s at the remaining $N-m$ positions (recall from Sec. \ref{sec:algo} that $m$ corresponds to the number of bits for the circular shift). The resulting value is shifted left by $N-m$ bits using a logarithmic bit shifter and \rev{temporarily} stored in \rev{register A}\footnote{\immodel~can perform two of such operations by using the register B to store the results, simultaneously to register A.}. At the same time, the data on the destination PE is shifted right by $m$ bits and saved in a spare row in the same subarray using the copy drivers. The value from \rev{register A} is moved to a second spare row in the destination PE, and an OR operation is performed between the values in these two spare rows to produce a circular right-shifted value, which is stored in a third spare $j^{th}$ row in the destination PE. Note that the original data of the source and destination PEs remain intact in $i^{th}$ row during the permutation.
 
 {\rev{\textbf{Round 2 (R2): }} In \rev{round 2}, the former \rev{source} PE becomes the new \rev{destination} PE, and \rev{the process described for round 1 repeats} until all PEs holding the HV have been used as destination PEs once. \rev{Afterwards}, all the permuted values stored in spare $j^{th}$ rows are copied to \rev{}{$i^{th}$ row of bundle segment of \immodel~to store the newly encoded HV.}
 
\rev{In Fig.~\ref{fig:Permutation}, we depict an example for the steps involved in the two-round permutation with IM-ODHD. The example performs a 2-bit circular shift (amount of 2 bits, to the right) on the string `ABCDEFGHIJKLMNOP', which results in `OPABCDEFGHIJKLMN'. The string is grouped into substrings of 4 characters and stored in four PEs, labeled as source (src) and destination (dest) PEs. A step-by-step explanation of the permutation with \immodel~is below:}

\begin{itemize}
    \item[] \rev{\textbf{R1-step (a), Fig.~\ref{fig:Permutation}(a):} Initially, the substrings in the source PEs are subjected to an in-memory AND operation with a `0011' mask.}
    \item[] \rev{\textbf{R1-step (b), Fig.~\ref{fig:Permutation}(b):} The masked substrings from Step (a) undergo a left shift, and the results get stored in registers A and B, placed near the CiM PEs.}
    \item[] \rev{\textbf{R1-step (c), Fig.~\ref{fig:Permutation}(c):} Simultaneous to step (a), the substrings in the destination PEs are subjected to an in-memory AND operation with a `1100' mask.}
    \item[] \rev{\textbf{R1-step (d), Fig.~\ref{fig:Permutation}(d):} Parallel to step (b), the masked substrings from Step (c) undergo a right shift, and are stored in the 1st spare row in the destination tiles.}
    \item[] \rev{\textbf{R1-step (e), Fig.~\ref{fig:Permutation}(e):} Substrings in registers A and B got moved to a 2nd spare row in the destination PEs.}
    \item[] \rev{\textbf{R1-step (f), Fig.~\ref{fig:Permutation}(f):} The results of an in-memory OR between the contents of the 1st and 2nd spare rows get stored in the 3rd spare row of the destination PEs, marking the end of the first round of permutation.}
    \item[] \rev{\textbf{R2, Fig.~\ref{fig:Permutation}(g):} Steps (a) through (f) of R1 repeat, with source PEs becoming destination PEs and vice-versa.}
\end{itemize}

\rev{Going through the steps (a) through (f) of round 1, and round 2, permutes the original string `ABCDEFGHIJKLMOP', resulting in `OPABCDEFGHIJKLMN'.}

 \begin{figure}[t]
      \centering
      {
          \includegraphics[width=0.81\columnwidth]{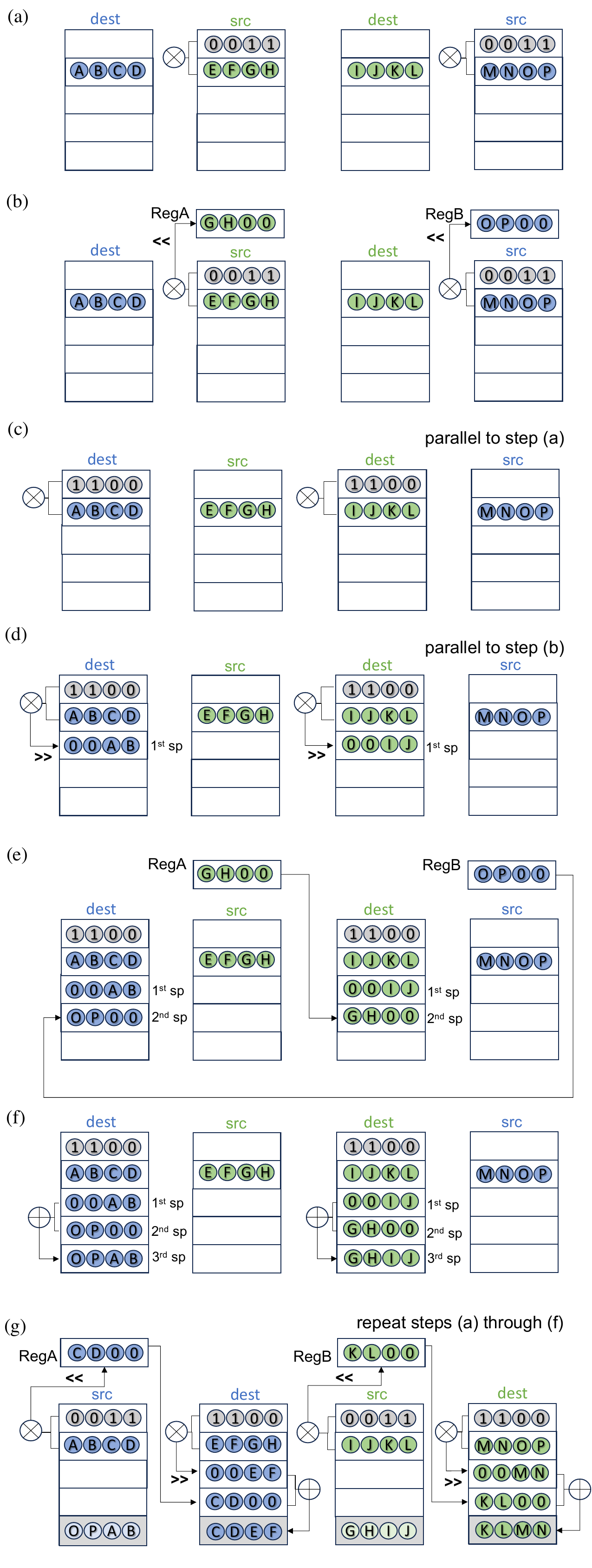}
      }
      \vspace{-3ex}
  	\caption{An example for the permutation with our CiM architecture; (a-f) corresponds to the steps of round 1; (g) depicts round 2.   }
   \vspace{-4ex}
  	\label{fig:Permutation}
  \end{figure}

 \textbf{Bundling on \immodel~}is implemented through the in-memory addition of the \rev{$m$} permuted HVs with the CSA. Before the in-memory addition, the permuted HVs are positioned in the \immodel~fabric so that the HV elements of the same column map to the same PE (but in different rows). \rev{For this, at first we copy the $1^{st}$ un-permuted feature HV $\vv{s_{f_0}}$ from the seed HV segment to the $1^{st}$ spare row of \immodel~fabric. The permuted HV for the second element of the original feature vector is duplicated to the next spare row. Once the positioning is done, the two HVs are added at a time with the CSA and the intermediate result is overwritten to the $1^{st}$ row. In the next round, the content of this row is added to the next permuted HV, until the bundling of all HVs is concluded. The result generates one encoded HV i.e. feature vector $\vv{F_n}$ per training dataset. To get the encoded feature HV of the next training sample we copy the $1^{st}$ un-permuted feature HV $\vv{s_{f_0}}$ from the seed HV segment to the $2^{nd}$ row of bundle segment of \immodel~fabric and repeat the operation for all of the training dataset in the consecutive rows.}
 %\rev{across $m \times D $ segment of \immodel
 %Before the in-memory addition, the permuted HVs are positioned in the \immodel~fabric so that the HV elements of the same column map to the same PE (but in different rows). \rev{For this, at first we copy the $1^{st}$ un-permuted feature HV $\vv{s_{f_0}}$ from the seed HV segment to the $1^{st}$ row of bundle segment and then keep copying the final permuted feature HV $\rho^{i-1}(\vv{s_{f_i}})$ from $j^{th}$ row to the $i^{th}$ row in bundle segment once the permutation is complete. Such positioning ensures that the HVs to be bundled are column-aligned for bitline-level computation at the CSA.}%, which is made possible through the use of the mat-level row/column decoders, as well as the source/destination registers in our architecture. Two permuted HVs are added at a time with the CSA, and the intermediate result is stored in a spare row. In the next round, the content of the spare row is added to the next permuted HV, until the bundling of all HVs is concluded. The result generates one encoded HV i.e. feature vector $\vv{F_n}$ per training dataset.}
 
 \subsubsection{Training in \immodel}

 Our CiM architecture generates the one-class HV ($\vv{H_{OC}}$) of the entire training set, i.e., all inlier samples, leveraging the bundling operation exactly as described in Sec. \ref{sec:hw_sw_encoding}.

\subsubsection{HW/SW Codesign for Threshold Calculation}

\label{sec:thresh_mod} 
The threshold calculation in Eq.~\ref{eqn:thres} uses mean and standard deviation, which requires division and a square root operation, which are not well supported by our proposed CiM architecture. Therefore, to make threshold calculation less computationally expensive to implement with our CiM architecture, we carry out three modifications to the algorithm proposed in Sec. \ref{sec:algo}. Namely, when running \model~on \immodel, we~(1) realize division with bit-shifts (enabled by logarithmic bit shifters), (2) modify the cosine similarity calculation, and (3) replace standard deviation with a more CiM-friendly mean absolute deviation (MAD) metric.

For (1), most CiM architectures (including ours) are not designed to efficiently support the division. In \immodel, division with our CiM hardware is approximated by shifting a binary value by $m$ bits to the right, which divides the value by $2^{m}$ and rounds down. The logarithmic bit shifters in \immodel~support a shift amount of 0-3 bits to the left or right. Therefore, divisions up to $2^3$ are possible, which are controlled by the shift mask (see Fig. \ref{fig:architecture}). Moreover, larger shift amounts (for larger divisors) are supported with multiple rounds of bit shifting. Since we need the division operation for calculating $\mu(S)$, we increase the training set such that the number of training samples equals a power of two value. This is done by copying random samples from the original training set without replacement. Doing so may increase the training time since the encoding phase has more samples to extract the information from. The impact of this is reflected in the training time presented in Sec. \ref{Time_Energy}.

%However, on the other hand, the number of samples that are identical in the training dataset is too small than that of the existing diverse samples, for which the~\model~does not become too familiar with those examples. Hence,~\model~does not overfit and can generalize to new, unseen data in the inference.

\begin{figure}[b]
     \centering
     {
         \includegraphics[width=1\columnwidth]{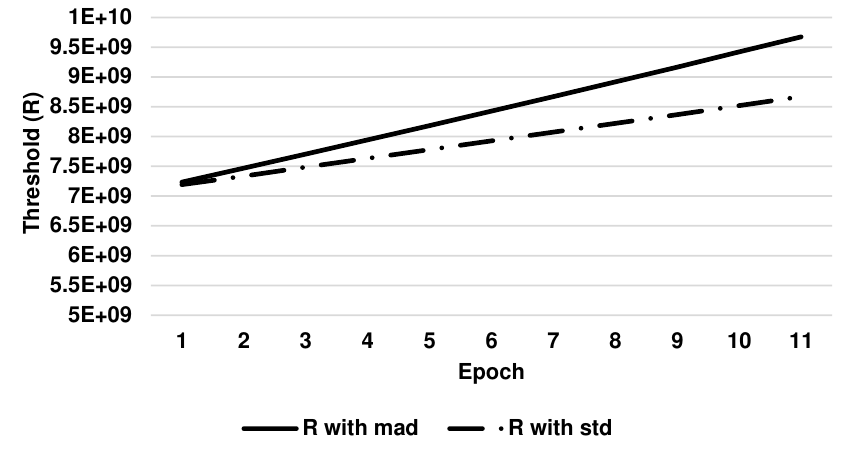}
     }
 	\caption{ Threshold trend on MNIST dataset with both standard deviation and mean absolute deviation metrics.}
 	\label{fig:thresh}
 \end{figure}
In the case of (2), while cosine similarity is used in Sec.~\ref{sec:algo} and in \cite{wang2022odhd}, our version of \immodel~uses only the numerator part of the cosine distance (see Eq. \ref{eqn:sim}) to make the architecture more amenable to CiM by eliminating square and square root operations \cite{Ranjan2019XMANNAC}. This is basically the binding operation or the dot product of two HVs, i.e., the one-class HV $\vv{H_{OC}}$ and the training HVs, which generates the similarity array $S$, followed by a sum (as shown in Eq.~\ref{eqn:sim_mod}).  \rev{This sum operation, essentially a pop-count (counting the number of 1s in a vector), is executed across several cycles. This operation hinges on accumulating partial sums using in-memory adders and bit shifters, key components of the subarrays in \immodel. Our evaluation incorporates this multi-cycle approach, which consists of the accumulation of these partial sums with the mentioned circuits.}

 \begin{equation}
 \small
     \delta(\vv{H_x}, \vv{H_y}) = \vv{H_x} \cdot \vv{H_y} = \sum_{i=1}^{d} h_{xi} \cdot h_{yi}
 \label{eqn:sim_mod}
 \end{equation}
 
Regarding (3), mean absolute deviation (MAD) is defined as the average absolute deviation of a set of values from their mean. MAD is calculated by finding the absolute difference between each data point and the mean (calculated in the previous step), summing these differences, and then dividing by the number of data points. All these operations (modular subtraction, addition, and division with bit shifts) are promptly supported by the components of our CiM architecture. Subtraction, for instance, can be performed as a 2`s complement subtraction where we first negate the subtrahend with a NOT operation, perform a local write of the result to the same subarray with a copy driver, and then finally perform an in-memory addition setting the carry of the first bit to 1. 

%Standard deviation and MAD are both indicators of how much the individual data points in a set differ from the central tendency. Since the absolute value function considers positive and negative deviations equally, is less susceptible to outliers, and accentuates extreme values in the data set, mean absolute deviation lays more attention on the extreme values in the data set. In the Fig.~\ref{fig:thresh} we can see the threshold increment per epoch for MNIST dataset from both the methods described in Sec.~\ref{sec:threshold} and Sec. ~\ref{sec:thresh_mod}. 

 Both standard deviation and MAD are measures of how much the data points in a set deviate from the mean. The absolute value function used in MAD treats positive and negative deviations equally, making it more robust to outliers and emphasizing extreme values in the dataset. As a result, MAD focuses more on the extreme values in the dataset than standard deviation does. Fig.~\ref{fig:thresh} shows the epoch-wise threshold increment for the MNIST dataset using both methods described in Sec.~\ref{sec:threshold} and Sec.~\ref{sec:thresh_mod}. We observe a higher threshold for MAD, with a difference that leads to the need for more epochs. Using the described methods, the mean value $\mu(S)$ and the mean absolute deviation $MAD(S)$ are computed for each similarity value in the array $S$ to calculate the threshold $R$.  The small modifications to the threshold estimation approach described in Sec.~\ref{sec:threshold} are reflected in Eq. \ref{eqn:thres_mod}. 
 \begin{equation}
     R = \mu(S) + 2*MAD(S)
     \label{eqn:thres_mod} 
 \end{equation}

% \subsubsection{Division in IMC-architecture}
% \label{subsec:Div}
% IMC architectures are not designed to perform arithmetic operations, such as division. They leverage from the bit shifting i.e. a bitwise operation that requires shifting a binary number's bits to a certain number of locations to the left or right. The operation shift logical right refers to halving and rounding down a number. Because it is done using bitwise operators that are supported by hardware and can be executed very quickly, bit shifting are especially effective in IMC. It can effectively split an integer by a power of two while doing division. For instance, we can move a number's bits one place to the right to divide it by 2, which is similar to dividing by $2^{1}$ \cite{cheng2019functional}. Since we need division operation for calculating $S_{mean}$ and $S_{mad}$ we can increase the training set such that number of training samples can be realized as $2^{N}$. In this work, it is done by copying random samples from the original training set without replacement.
% Though this might increase the training time since the encoding phase has more sample to extract the information from but it makes the model less likely to overfit providing more data to learn from.

\subsubsection{HW/SW Codesign for Fine-Tuning}

\label{sec:finetuning_hw_sw} 

As detailed in Sec.~\ref{sec:algo}, fine-tuning is used to ensure that all training HVs will be correctly identified as inliers. During each fine-tuning epoch, we use all the training samples\rev{, previously encoded in the encoding phase of \immodel}, and for each individual training sample, we calculate the similarity between its HV and the one-class HV $\vv{H_{OC}}$ using \review{Eq. \ref{eqn:sim_mod}}. All the misclassified inliers are then updated into the one-class HV $\vv{H_{OC}}$ exploiting the in-memory addition with the CSA as described in Sec. \ref{sec:background_CiMcircuits}. \rev{Once fine-tuning is accomplished, we no longer need to store the encoded training samples and only store the one-class HV $\vv{H_{OC}}$ along with the seed HVs that are used in Sec.~\ref{sec:detection_hw_sw}.}

%While cosine similarity is used in Sec. ~\ref{sec:algo} and in \cite{wang2022odhd}, our version of \immodel~uses only the numerator part of the cosine distance (see Eq. \ref{eqn:sim}) to make the architecture more amenable to CiM by eliminating square and square root operations (similar to what has been done in \cite{Ranjan2019XMANNAC}). This is basically the binding operation or the dot product of two HVs, i.e., the one-class HV $\vv{HV_{OC}}$ and the training HVs, which generates the similarity array $S$, followed by a sum (as shown in Eq.~\ref{eqn:sim_mod}).  
% \begin{equation}
% \small
%     \delta(\vv{H_x}, \vv{H_y}) = \vv{H_x} \cdot \vv{H_y} = \sum_{i=1}^{d} h_{xi} \cdot h_{yi}
% \label{eqn:sim_mod}
% \end{equation}

%\textbf{Binding with our \immodel~fabric} can be performed directly by the CSA in the form of a bitwise AND operation. 
 
 \subsubsection{Outlier Detection in \immodel}

\label{sec:detection_hw_sw} 

Once we have trained the $\vv{H_{OC}}$ and established the threshold $R$ using the described CiM-friendly method, we deploy \model~to detect outliers in unseen samples without knowledge of their labels. During the outlier detection phase (i.e. the inference/test phase), we encode the testing sample $q$ into a query HV using the same encoding process described in Sec. \ref{sec:hw_sw_encoding} and the same seed HVs. We then calculate the similarity between the query HV and the one-class $\vv{H_{OC}}$ using the method described in Sec. \ref{sec:finetuning_hw_sw}. If the similarity is lower than the predetermined threshold, sample $q$ is classified as an outlier.

\section{Evaluation}
\label{sec:exp}
In this section, we evaluate the performance of \model ~on six datasets and compare the CiM, CPU, and GPU-based implementations of \model ~with four baseline methods. 

\subsection{Experimental Setup}
\label{sec:setup}
Herewith, we discuss the experimental setup for our software and hardware-level evaluations.

\subsubsection{Software Evaluation}
We evaluate the performance of \model ~on six datasets selected from the Outlier Detection Datasets (ODDS) Library~\cite{Rayana2016} spanning multiple application domains such as medical diagnosis and wireless communication. These datasets are Wisconsin-Breast Cancer (Diagnostics) dataset (WBC), Mammography (MAMMO), MNIST, Cardiotocography (CARDIO), lymphography (LYMPHO), and Landsat Satellite (SATI2). These datasets are widely used as benchmarks in existing outlier detection studies~\cite{zimek2013subsampling, sathe2016lodes}. Each dataset contains a certain number of outliers specified by the ODDS library, e.g., the WBC dataset has 21 outliers. The testing dataset is mixed inliers and outliers, e.g., $25\%-75\%$ outlier - inlier mix. More details of these datasets can be found on the (ODDS) Library website~\cite{Rayana2016}. 

We repeat the experiments independently 10 times and report the average performance. We also present error bars as shown in Fig.~\ref{fig:comp} to illustrate the performance variations due to the randomness in different learning methods. 

We compare our modified \model~implemented on \immodel~discussed in Sec. \ref{sec:hwsw_codesign}~with the original \model~algorithm proposed in \cite{wang2022odhd} and discussed in Sec. \ref{sec:algo}, as well as with the following four baseline outlier detection methods: 
% neural network-based autoencoder~\cite{he2020exploring}, isolation forests~\cite{liu2008isolation}, OCSVM~\cite{7726152}, and HDAD~\cite{wang2021brief}. The implementation details are as follows:  
\begin{itemize}
    \item \textbf{Autoencoder}: Autoencoder is an emerging unsupervised learning outlier detection approach based on a neural network. In this paper, we use the same autoencoder architecture as~\cite{he2020exploring}.
    \item \textbf{Isolation Forest}: Isolation Forest is an ensemble model of isolation trees, which uses the path length of each sample to detect outliers. In this paper, we establish an isolation forest model using the same configuration as~\cite{liu2008isolation}.
    \item \textbf{OCSVM}: OCSVM attempts to separate outliers from the inliers with the maximum margin. We have a grid search for an appropriate set of hyper-parameters such as kernel functions and the value of gamma to fine-tune the OCSVM model following~\cite{wang2018hyperparameter}.
    \item \textbf{HDAD}: HDAD follows similar principles of autoencoder; it first ``reconstruct'' the input samples and then detects anomalies based on reconstruction error. We use the same architecture of ~\cite{wang2021brief}.
\end{itemize}

We implement \model~and the four baseline methods in Python and perform our experiments on a desktop with an i7-7700 CPU, 12 GB RAM, and an NVIDIA P1000 GPU with 4 GB onboard memory. \rev{We implement the GPU version ODHD based on Pytorch and use the HWiNFO tool~\cite{hwinfo} to measure energy consumption. HWiNFO is a commercial tool for monitoring hardware circumstances and has been utilized in previous work ~\cite{mo2023haac, maxwell2021using}. }

\rev{Unlike traditional DNN operations such as the Conv2D layer, the GPU version of ODHD does not have a specialized data flow or CUDA optimization. In the GPU implementation, the most time-consuming part is data transfer between the CPU and GPU memory. Since the HV are high-dimension vectors, the GPU acceleration can be slowed down by the data transfer. According to our experimental results, the GPU version of ODHD provides $\sim$2--2.5$\times$ time efficiency compared with the CPU version of ODHD, which is consistent with the results presented in torchHD~\cite{heddes2022torchhd}.} 

To comprehensively assess the algorithm-level performance of \model, we use three metrics: accuracy (ACC), F1 score (F1), and Area under ROC curve ROC-AUC (AUC). Note that while accuracy is widely used and easy to understand, an outlier detection dataset may be significantly imbalanced. Hence, accuracy may not precisely reveal the performance of outlier detectors. Therefore, we also use ROC-AUC, which is widely used for outlier detection as it can accurately represent the tradeoff between true positive and false positive~\cite{wang2020further}. Meanwhile, the F1 score is also a widely-used metric in binary classification which can comprehensively indicate the tradeoff between precision and recall~\cite{everingham2010pascal}.

\subsubsection{Hardware Evaluation}

\begin{table}[t]
   \centering
   \caption{Parameters used in our evaluation of the CiM mat}
    % \vspace{-10pt}
\begin{tabular}{@{}lllll@{}}
\toprule
Parameter & P  & Q  & M    & N    \\ \midrule
\review{\textbf{PE (L)}: Large PE }     & 16 & 16 & 1024 & 1024 \\ 
\review{\textbf{PE (M)}: Medium PE}      & 32 & 32 & 512 & 512 \\ 
\review{\textbf{PE (S)}: Small PE}      & 64 & 64 & 256 & 256 \\ \bottomrule
\end{tabular}
\label{tab:cimparam}
\end{table}

\begin{table}[t]
 \small
   \centering
   \caption{Latency (ns) and energy (nJ) for IM operations, \review{with respect to the architectures defined in Table~\ref{tab:cimparam}.}}
    % \vspace{-10pt}
\resizebox{\columnwidth}{!}{%
\begin{tabular}{@{}ccccccc@{}}
\toprule
{\color[HTML]{000000} }                                                          & \multicolumn{3}{c}{{\color[HTML]{000000} Latency (ns)}}                                                                                                                                                                           & \multicolumn{3}{c}{{\color[HTML]{000000} Energy (nJ)}}                                                                                                                                                                            \\
\multirow{-2}{*}{{\color[HTML]{000000} Operation}}                               & {\color[HTML]{000000} \begin{tabular}[c]{@{}c@{}}\review{PE (L)} \end{tabular}} & {\color[HTML]{000000} \begin{tabular}[c]{@{}c@{}}\review{PE (M)}\end{tabular}} & {\color[HTML]{000000} \begin{tabular}[c]{@{}c@{}}\review{PE (S)}\end{tabular}} & {\color[HTML]{000000} \begin{tabular}[c]{@{}c@{}}\review{PE (L)}\end{tabular}} & {\color[HTML]{000000} \begin{tabular}[c]{@{}c@{}}\review{PE (M)}\end{tabular}} & {\color[HTML]{000000} \begin{tabular}[c]{@{}c@{}}\review{PE (S)}\end{tabular}} \\ \midrule
%\multirow{-2}{*}{{\color[HTML]{000000} Operation}}                               & {\color[HTML]{000000} \begin{tabular}[c]{@{}c@{}}Design\\ 1\end{tabular}} & {\color[HTML]{000000} \begin{tabular}[c]{@{}c@{}}Design\\ 2\end{tabular}} & {\color[HTML]{000000} \begin{tabular}[c]{@{}c@{}}Design\\ 3\end{tabular}} & {\color[HTML]{000000} \begin{tabular}[c]{@{}c@{}}Design\\ 1\end{tabular}} & {\color[HTML]{000000} \begin{tabular}[c]{@{}c@{}}Design\\ 2\end{tabular}} & {\color[HTML]{000000} \begin{tabular}[c]{@{}c@{}}Design\\ 3\end{tabular}} \\ \midrule
{\color[HTML]{000000} Read/NOT}                                                  & {\color[HTML]{000000} 5.24}                                               & {\color[HTML]{000000} 2.64}                                               & {\color[HTML]{000000} 1.42}                                               & {\color[HTML]{000000} 17.36}                                              & {\color[HTML]{000000} 5.51}                                               & {\color[HTML]{000000} 1.66}                                               \\
{\color[HTML]{000000} AND/OR}                                                    & {\color[HTML]{000000} 5.28}                                               & {\color[HTML]{000000} 2.68}                                               & {\color[HTML]{000000} 1.48}                                               & {\color[HTML]{000000} 18.44}                                              & {\color[HTML]{000000} 18.40}                                              & {\color[HTML]{000000} 2.50}                                               \\
{\color[HTML]{000000} \begin{tabular}[c]{@{}c@{}}Pointwise\\ Mult.\end{tabular}} & {\color[HTML]{000000} 5.28}                                               & {\color[HTML]{000000} 2.68}                                               & {\color[HTML]{000000} 1.48}                                               & {\color[HTML]{000000} 18.44}                                              & {\color[HTML]{000000} 18.40}                                              & {\color[HTML]{000000} 2.50}                                               \\
{\color[HTML]{000000} Write}                                                     & {\color[HTML]{000000} 5.08}                                               & {\color[HTML]{000000} 2.46}                                               & {\color[HTML]{000000} 1.26}                                               & {\color[HTML]{000000} 14.58}                                              & {\color[HTML]{000000} 6.78}                                               & {\color[HTML]{000000} 0.96}                                               \\
{\color[HTML]{000000} Add}                                                       & {\color[HTML]{000000} 12.87}                                              & {\color[HTML]{000000} 10.20}                                              & {\color[HTML]{000000} 9.04}                                               & {\color[HTML]{000000} 19.97}                                              & {\color[HTML]{000000} 96.30}                                              & {\color[HTML]{000000} 47.30}                                              \\
{\color[HTML]{000000} Sub}                                                       & {\color[HTML]{000000} 17.96}                                              & {\color[HTML]{000000} 12.70}                                              & {\color[HTML]{000000} 10.30}                                              & {\color[HTML]{000000} 21.43}                                              & {\color[HTML]{000000} 103.08}                                             & {\color[HTML]{000000} 48.21}                                              \\
{\color[HTML]{000000} Shift}                                                     & {\color[HTML]{000000} 5.24}                                               & {\color[HTML]{000000} 2.64}                                               & {\color[HTML]{000000} 1.42}                                               & {\color[HTML]{000000} 17.36}                                              & {\color[HTML]{000000} 5.51}                                               & {\color[HTML]{000000} 1.66}                                               \\
{\color[HTML]{000000} Permut*}                                              & {\color[HTML]{000000} 36.13}                                              & {\color[HTML]{000000} 17.80}                                              & {\color[HTML]{000000} 9.40}                                               & {\color[HTML]{000000} 93.58}                                              & {\color[HTML]{000000} 69.50}                                              & {\color[HTML]{000000} 10.50}                                              \\ \bottomrule
\end{tabular}%
}
   \label{tab:arraylevel}
   \begin{flushright}
\scriptsize
\rev{\textbf{*} For permutation, $\sim$56.2\% of the time ($\sim$37.7\% of the energy) is spent on operations between the PEs and the registers, while the rest of the time (energy) is spent on operations performed within the PEs. } 
\vspace{-3ex}
\end{flushright}

 \end{table}

 \begin{figure*}[t]
     \centering
          \includegraphics[width = 0.98\columnwidth]{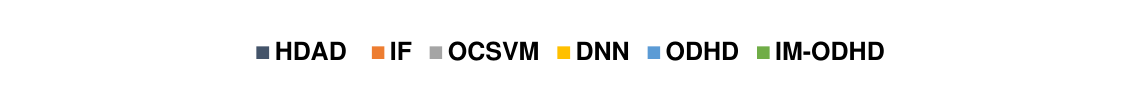}
 	\vspace{-5pt}
 	
     \subfigure[ACC]{
         \includegraphics[width=0.63\columnwidth]{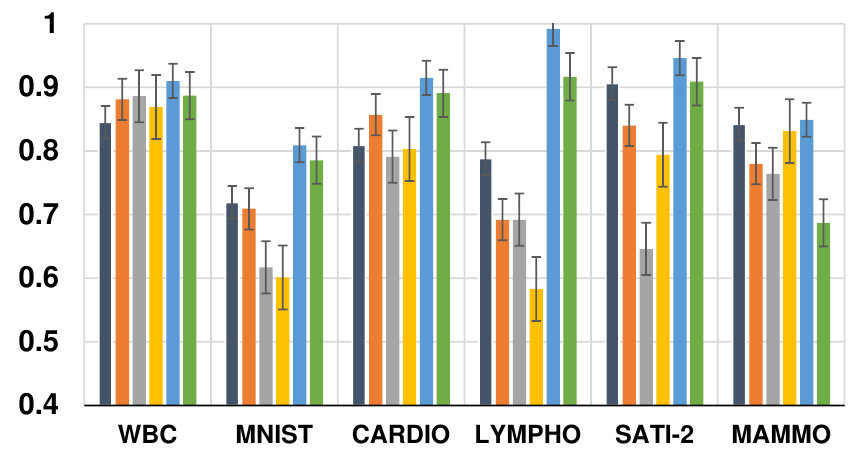}
     }
     \hspace{5pt}
     \subfigure[F1]{
         \includegraphics[width=0.63\columnwidth]{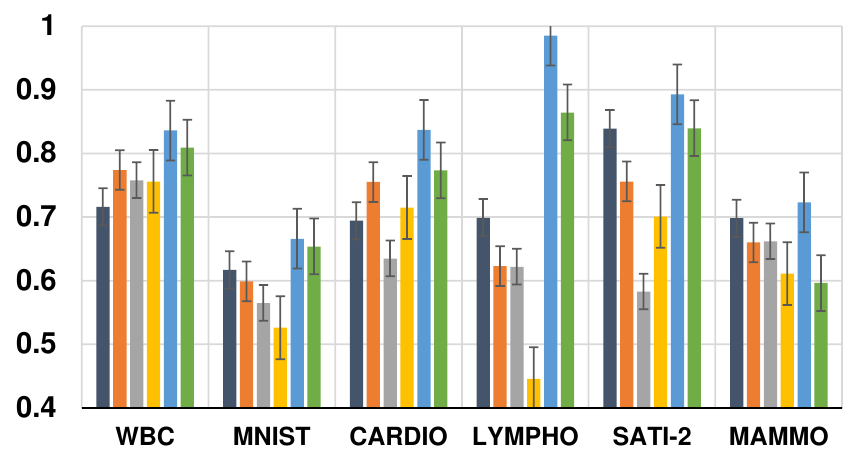}
     }
     \hspace{5pt}
     \subfigure[AUC]{
         \includegraphics[width=0.63\columnwidth]{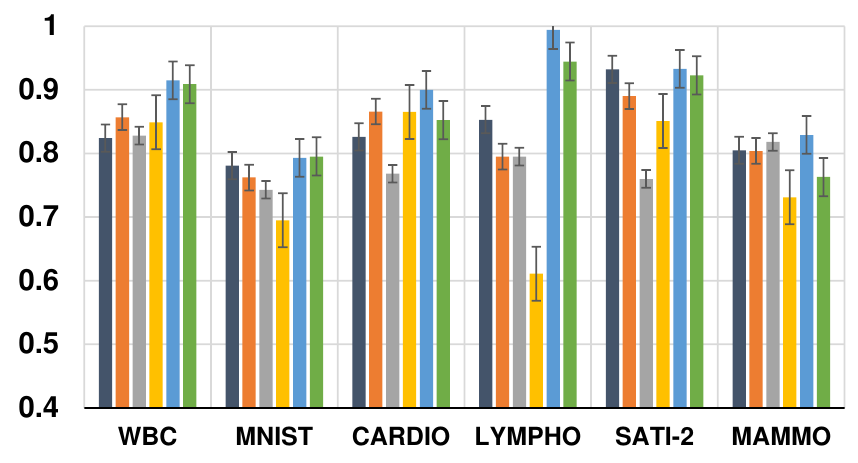}
     }

 	\caption{Comparison between \model~, \immodel~and four baseline methods based on three metrics, ACC, F1 and ROC. }
 	\label{fig:comp}
 \end{figure*}

 \begin{figure*}[t]
     \centering

     \subfigure[ACC]{
         \includegraphics[width=0.63\columnwidth]{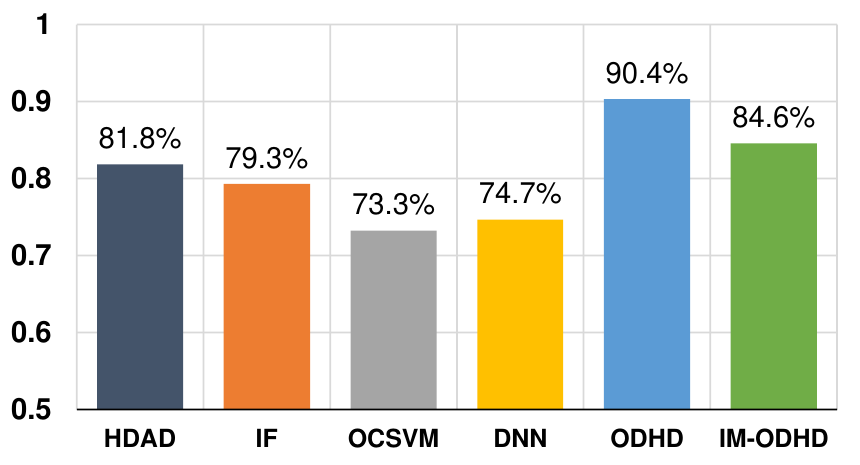}
     }
     \hspace{5pt}
     \subfigure[F1]{
         \includegraphics[width=0.63\columnwidth]{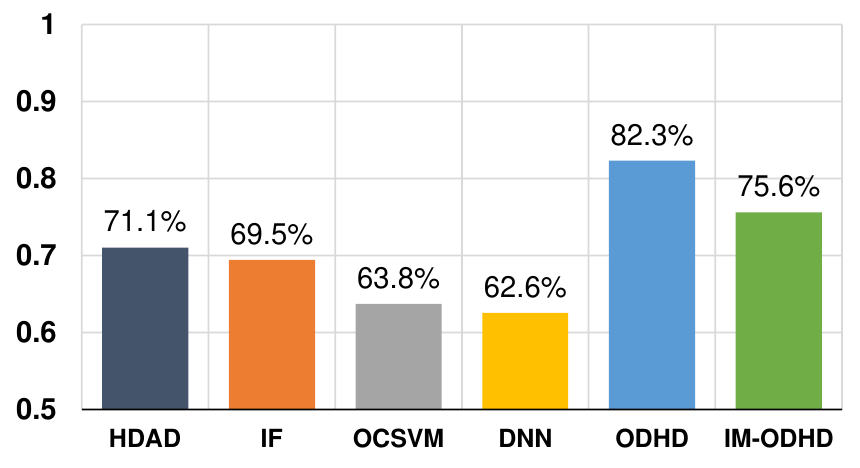}
     }
     \hspace{5pt}
     \subfigure[AUC]{
         \includegraphics[width=0.63\columnwidth]{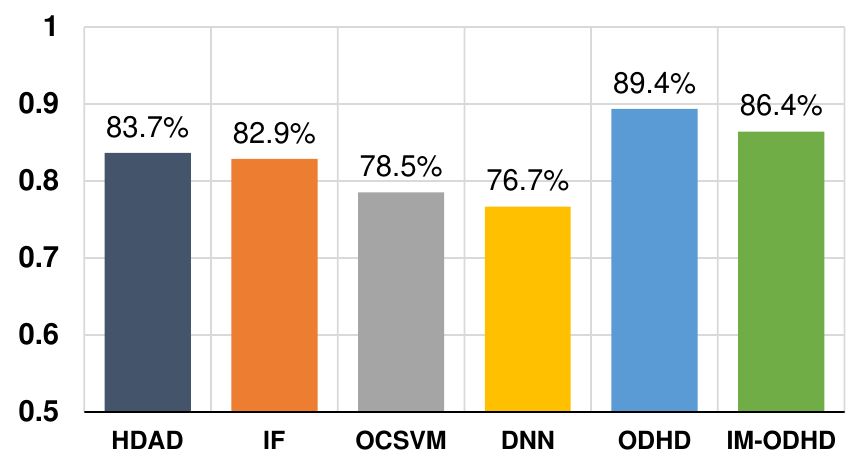}
     }

 	\caption{ Average performance of different models over six datasets.}
 	\label{fig:comp_avg}
 \end{figure*}

Besides the accuracy, F1, and AUC, which are used to evaluate the algorithm-level performance of \model, we measure the runtime and energy of \model~based on different platforms, i.e., CPU, GPU, and CiM, to capture performance tradeoffs. For the CPU version of \model, we measure the training and testing runtime of the application running on a desktop with an i7-7700 CPU, 12 GB RAM, and an NVIDIA P1000 GPU with 4 GB on-board memory. 

For the CiM implementation of \model, which we refer to as \immodel, we simulated the SRAM-based CiM mat of Fig. \ref{fig:architecture} using Destiny \cite{poremba2015destiny}, a tool for modeling emerging 2D and 3D NVM and SRAM caches, which was extended with the customized peripheral circuits employed by \immodel~(i.e., customized sense amplifiers, logarithmic bit shifters, an extra wordline decoder, and copy drivers). In our evaluation, we employ the CMOS Predictive Technology Model (PTM) from \cite{zhao2007predictive}, specifically designed for a 45nm technology node \rev{to simulate CiM circuits. Furthermore, our evaluation of \immodel~accounted for all the components shown in Fig.~\ref{fig:architecture} in addition to those within each subarray. The registers, mat-level decoders, and communication network were implemented using Verilog and synthesized with Cadence Encounter RTL Compiler v14.10, using the NanGate 45nm open-cell library \cite{knudsen2008nangate}.}

\rev{To accommodate a wide range of datasets with our SRAM-based CiM architecture, we must carefully select values for P, Q, M, and N. We conduct a design space exploration for the \immodel~mat parameters, as outlined in Table~\ref{tab:cimparam}. Table~\ref{tab:arraylevel} summarizes the latency and energy consumption of various in-memory operations across the three simulated design configurations from Table~\ref{tab:cimparam}.}

\rev{Notably, PE (S) exhibits shorter latency, along with reduced energy consumption per PE. However, PE (S) also results in an increased number of dedicated PEs, which necessitates an expanded global address line and a larger amount of memory peripherals, leading to a potential area disadvantage with respect to PE (M) and PE (L). In contrast, PE (L) results in less peripheral circuitry and it is more area efficient, at the same time providing satisfactory latency and energy. To establish a lower-bound for the performance of \immodel, we employ \review{PE (L)} throughout our latency and energy evaluation (Sec. \ref{Time_Energy}).}

\subsection{Performance Evaluation for \model}

As shown in Fig.~\ref{fig:comp}, we compare the performance of \model~and \immodel~with the baseline methods on six datasets for the three metrics \textit{with error bar}. \immodel~performs better than the baseline models in all but one case (as will be discussed in Sec. \ref{sec:hw_sw_codesign_eval}). On the other hand, \model~is able to consistently outperform the four baseline methods \textbf{on every dataset for every metric}. 
%On the other hand, the CPU version of \immodel~(which we refer to as \model) is able to consistently outperform the four baseline methods \textbf{on every dataset for every metric}. 

First, for F1, the average F1 of \model~is 82.3\% on all datasets, representing an improvement of 18.5\% over OCSVM, 12.8\% over isolation forest, 15.8\% over HDAD and 19.8\% over autoencoder.
For AUC, the average AUC of \model~is 89.4\%, representing an improvement of 10.9\% over OCSVM, 6.5\% over isolation forest, 6.8\% over HDAD, and 12.7\% over autoencoder.

Second, while \model~has a certain level of fluctuation (error bars) in different runs (just like all the other models), we can observe that even the low end of \model~is higher than the high end of any baseline method, representing the robustness of the performance of \model. 

Third, while the performance of different methods varies with different datasets, \model~shows better stability compared to other methods. For example, for ACC, the lowest ACC of \model~is over 80\%, while the lowest ACC are about 60\%, 70\%, 60\% and 70\% for OCSVM, Isolation Forest, Autoencoder and HDAD, respectively. A similar phenomenon can also be seen in F1 and AUC. 

Last but not least, in certain datasets, e.g., LYMPHO, all baseline methods significantly underperform while \model~maintains a high accuracy close to 100\%. The reason is possibly related to the fact that the lymphography data are relatively small so that the baseline methods cannot converge to a proper point; however, \model~is able to learn useful information even from a small amount of data. Similar advantages of HDC have been observed in various supervised classification studies for biomedical datasets that are often small~\cite{burrello2018one}. 

 \begin{table*}[t]
 \small
   \centering
   \caption{\rev{Latency/energy breakdown of training with \immodel~\textcolor{black}{using PE (L)}}}
\resizebox{2\columnwidth}{!}{%
\begin{tabular}{@{}ccccccccc@{}}
\toprule
{\color[HTML]{000000} } &
\multicolumn{3}{c}{{\color[HTML]{000000} \textbf{Encoding: Permutation}}} &
  {\color[HTML]{000000} \textbf{Enc: Bundling}} &
  {\color[HTML]{000000} \textbf{Bndl+Thr+Tun}} &
  {\color[HTML]{000000} } \\
\multirow{-2}{*}{{\color[HTML]{000000} \textbf{Dataset}}} &
  {\color[HTML]{000000} \begin{tabular}[c]{@{}c@{}}IM Ops \\w. PEs ($\mu$s/$\mu$J)\end{tabular}} &
  {\color[HTML]{000000} \begin{tabular}[c]{@{}c@{}}PE-to-REG \\ Comm. ($\mu$s/$\mu$J)\end{tabular}} &
  {\color[HTML]{000000} \begin{tabular}[c]{@{}c@{}}REG-to-PE \\ Comm. ($\mu$s/$\mu$J)\end{tabular}} &
  {\color[HTML]{000000} \begin{tabular}[c]{@{}c@{}}IM Ops \\w.  PEs ($\mu$s/$\mu$J)\end{tabular}} &
  {\color[HTML]{000000} \begin{tabular}[c]{@{}c@{}}IM Ops \\w.  PEs ($\mu$s/$\mu$J)\end{tabular}} &
  {\color[HTML]{000000} \begin{tabular}[c]{@{}c@{}}Total \\ ($\mu$s/$\mu$J)\end{tabular}} \\ \midrule
{\color[HTML]{000000} WBC} &
  {\color[HTML]{000000} 307.6/980.8} &
  {\color[HTML]{000000} 153.3/474.3} &
  {\color[HTML]{000000} 75.5/216.5} &
  {\color[HTML]{000000} 17.2/205.8} &
  {\color[HTML]{000000} \review{91.1/1039.6}} &
  {\color[HTML]{000000} \review{644.8/2917.0}} \\
{\color[HTML]{000000} MNIST} &
  {\color[HTML]{000000} 16802.5/53570.3} &
  {\color[HTML]{000000} 8374.5/25908.5} &
  {\color[HTML]{000000} 4121.5/11826.9} &
  {\color[HTML]{000000} 919.2/10973.4} &
  {\color[HTML]{000000} \review{1450.8/16548.4}} &
  {\color[HTML]{000000} \review{31668.5/118827.5}} \\
{\color[HTML]{000000} CARDIO} &
  {\color[HTML]{000000} 848.6/2705.6} &
  {\color[HTML]{000000} 423.0/1308.5} &
  {\color[HTML]{000000} 208.2/597.3} &
  {\color[HTML]{000000} 48.3/576.1} &
  {\color[HTML]{000000} \review{363.0/4140.0}} &
  {\color[HTML]{000000} \review{1890.9/9327.5}} \\
{\color[HTML]{000000} LYMPHO} &
  {\color[HTML]{000000} 45.1/143.7} &
  {\color[HTML]{000000} 22.5/69.5} &
  {\color[HTML]{000000} 11.1/31.7} &
  {\color[HTML]{000000} 2.6/30.9} &
  {\color[HTML]{000000} \review{22.8/260.0}} &
  {\color[HTML]{000000} \review{104.0/535.8}} \\
{\color[HTML]{000000} SATI} &
  {\color[HTML]{000000} 5940.3/18939.0} &
  {\color[HTML]{000000} 2960.7/9159.6} &
  {\color[HTML]{000000} 1457.1/4181.2} &
  {\color[HTML]{000000} 330.9/3950.4} &
  {\color[HTML]{000000} \review{1459.1/16647.6}} &
  {\color[HTML]{000000} \review{12148.1/52877.8}} \\
{\color[HTML]{000000} MAMMO} &
  {\color[HTML]{000000} 1697.2/5411.1} &
  {\color[HTML]{000000} 845.9/2617.0} &
  {\color[HTML]{000000} 416.3/1194.6} &
  {\color[HTML]{000000} 110.3/1316.8} &
  {\color[HTML]{000000} \review{2903.8/33123.7}} &
  {\color[HTML]{000000} \review{5973.5/43663.3}} \\ \bottomrule
\end{tabular}%
}
   \label{tab:imodhd_breakdown_training}
 \end{table*}

\begin{table*}[]
 \small
   \centering
   \caption{\rev{Latency/energy breakdown of testing with \immodel~\textcolor{black}{using PE (L)}}}
\resizebox{2\columnwidth}{!}{%
\begin{tabular}{@{}ccccccc@{}}
\toprule
{\color[HTML]{000000} } &
  \multicolumn{3}{c}{{\color[HTML]{000000} \textbf{Encoding: Permutation}}} &
  {\color[HTML]{000000} \textbf{Enc: Bundling}} &
  {\color[HTML]{000000} \textbf{Outlier Detection}} &
  {\color[HTML]{000000} } \\
\multirow{-2}{*}{{\color[HTML]{000000} \textbf{Dataset}}} &
  {\color[HTML]{000000} \begin{tabular}[c]{@{}c@{}}IM Ops \\w. PEs ($\mu$s/$\mu$J)\end{tabular}} &
  {\color[HTML]{000000} \begin{tabular}[c]{@{}c@{}}PE-to-REG \\ Comm. ($\mu$s/$\mu$J)\end{tabular}} &
  {\color[HTML]{000000} \begin{tabular}[c]{@{}c@{}}REG-to-PE \\ Comm. ($\mu$s/$\mu$J)\end{tabular}} &
  {\color[HTML]{000000} \begin{tabular}[c]{@{}c@{}}IM Ops \\w.  PEs ($\mu$s/$\mu$J)\end{tabular}} &
  {\color[HTML]{000000} \begin{tabular}[c]{@{}c@{}}IM Ops \\w.  PEs ($\mu$s/$\mu$J)\end{tabular}} &
  {\color[HTML]{000000} \begin{tabular}[c]{@{}c@{}}Total \\ ($\mu$s/$\mu$J)\end{tabular}} \\ \midrule
{\color[HTML]{000000} WBC} &
  {\color[HTML]{000000} 50.7/161.7} &
  {\color[HTML]{000000} 25.3/78.2} &
  {\color[HTML]{000000} 12.4/35.7} &
  {\color[HTML]{000000} 2.8/33.9} &
  {\color[HTML]{000000} \review{1.5/18.7}} &
  {\color[HTML]{000000} \review{92.8/328.1}} \\
{\color[HTML]{000000} MNIST} &
  {\color[HTML]{000000} 5743.0/18310.2} &
  {\color[HTML]{000000} 2862.4/8855.4} &
  {\color[HTML]{000000} 1408.7/4042.4} &
  {\color[HTML]{000000} 314.2/3750.7} &
  {\color[HTML]{000000} \review{43.6/544.2}} &
  {\color[HTML]{000000} \review{10371.9/35502.9}} \\
{\color[HTML]{000000} CARDIO} &
  {\color[HTML]{000000} 291.7/930.0} &
  {\color[HTML]{000000} 145.4/449.4} &
  {\color[HTML]{000000} 71.6/205.3} &
  {\color[HTML]{000000} 16.6/198.0} &
  {\color[HTML]{000000} \review{12.0/136.8}} &
  {\color[HTML]{000000} \review{537.2/1920.0}} \\
{\color[HTML]{000000} LYMPHO} &
  {\color[HTML]{000000} 8.6/27.5} &
  {\color[HTML]{000000} 4.3/13.3} &
  {\color[HTML]{000000} 2.1/6.1} &
  {\color[HTML]{000000} 0.5/5.8} &
  {\color[HTML]{000000} \review{0.5/6.2}} &
  {\color[HTML]{000000} \review{16.1/58.8}} \\
{\color[HTML]{000000} SATI} &
  {\color[HTML]{000000} 206.2/657.4} &
  {\color[HTML]{000000} 102.8/317.9} &
  {\color[HTML]{000000} 50.6/145.1} &
  {\color[HTML]{000000} 11.5/137.0} &
  {\color[HTML]{000000} \review{4.9/56.0}} &
  {\color[HTML]{000000} \review{375.9/1313.4}} \\
{\color[HTML]{000000} MAMMO} &
  {\color[HTML]{000000} 107.7/343.5} &
  {\color[HTML]{000000} 53.7/166.1} &
  {\color[HTML]{000000} 26.4/75.8} &
  {\color[HTML]{000000} 7.0/83.6} &
  {\color[HTML]{000000} \review{17.7/202.1}} &
  {\color[HTML]{000000} \review{212.5/871.2}} \\ \bottomrule
\end{tabular}%
}
\label{tab:imodhd_breakdown_testing}
\end{table*}
 
\subsection{Performance Evaluation for \immodel}
\label{sec:hw_sw_codesign_eval}

From Fig.~\ref{fig:comp} it is evident that \immodel~shows modest accuracy loss with respect to \model~due to algorithm-level modification of \model. To better understand the extent of this accuracy loss, we analyze the average results for all datasets (presented in Fig. \ref{fig:comp_avg}).

For ACC, the average ACC of \model~is 90.4\% on all datasets, representing an improvement of 17.1\% over OCSVM, 11.1\% over isolation forest, 10.5\% over HDAD and 15.7\% over autoencoder. The average ACC, F1 score of \immodel~degrades 6.37\%, 8.1\% respectively compared to \model. The most important metric for outlier detection, ROC-AUC degrades only 3.3\% i.e., \immodel~can still satisfactorily represent the tradeoff between true positive and false positive. It is important to highlight that, even though these metrics do not surpass the software level accuracy for \model, \immodel~still shows better performance in terms of the average accuracy i.e. 3.4\% over HDAD, 6.6\% over isolation forest, 15.4\% over OCSVM, 13.3\% over DNN, the average F1 score i.e. 3.4\% over HDAD, 6.6\% over isolation forest, 15.4\% over OCSVM, 13.3\% over DNN and the average ROC i.e. 3.4\% over HDAD, 6.6\% over isolation forest, 15.4\% over OCSVM, 13.3\% over DNN. 

We observe that the performance of \immodel~is heavily dependent on how well the high variability feature HVs capture the diversity of each sample during the training stage. The results show that \immodel~does not perform as well with the mammography dataset as it does on other datasets, achieving an accuracy of 68.7\% and an F1-score of 59.6\%, which is lower than other existing models. However, \immodel~exhibits a 3.4\% improvement in the ROC-AUC metric compared to the baseline model DNN. Therefore, although \immodel~can achieve high accuracy with small datasets like lympho, its performance is highly dependent on the ability of the HVs to interpret the data from the features. With an increase in the number of features in the dataset, the accuracy of \immodel~approaches the software level (i.e., \model) accuracy. For instance, with the MNIST dataset, which has the highest number of features among all the six datasets, the accuracy of \immodel~is only 1\% less accurate than \model`s accuracy.

\subsection{Latency and Energy Evaluation}
\label{Time_Energy}

\begin{table*}[t]
 \small
   \centering
   \caption{Execution time (ms) of different outlier detection models over six datasets (training time/testing time)}
    % \vspace{-10pt}
     \resizebox{2\columnwidth}{!}{
     \begin{tabular}{ccccccc}
     
     \toprule
     {}       & {OCSVM}    & {Isolation Forest} & {HDAD} & {ODHD} & {ODHD(GPU)} & {\immodel} \\
     \midrule
     {WBC}    & 3.000/2.000 & 112.0/30.00  &412.0/198.0   & 399.0/112.0  & 187.0/62.00 & \review{0.645/0.093}  \\
     {MNIST}  &925.0/782.0 & 355.0/198.0 &20773/25662 & 18024/9631  & 8872/6159   & \review{31.67/10.372} \\
     {CARDIO} &31.00/45.00 & 115.0/42.00 &1134/1248   & 1212/615.0  & 511.0/442.0 & \review{1.89/0.537} \\
     {LYMPHO} &1.000/1.000 & 111.0/28.00 &169.0/53.00 & 119.0/20.00 & 45.00/12.00 &  \review{0.104/0.016} \\
     {SATI2}  &965.0/84.00 & 211.0/36.00 &7205/704.0  & 9109/549.0  & 4194/267.0  & \review{12.148/0.376} \\
     {MAMMO}  & 1942/478.0 & 151.0/44.00 &7474/1129   & 5644/555.0  & 2215/382.0  & \review{5.974/0.213}  \\
     \bottomrule
   
     \end{tabular}}
   \label{tab:execution_time}
 \end{table*}
 
\begin{table*}[t]
\small
% \vspace{10pt}
  \centering
  \caption{Energy (mJ) Comparison \model~vs. \immodel}
  \resizebox{1.5\columnwidth}{!}{
    \hspace{-10pt}
    \begin{tabular}{ccc}
    \hline
    \toprule

          {} & {ODHD (GPU)}                          & {\immodel} \\

   {Dataset} & {Training Energy}~~~ {Testing Energy} & {Training Energy}~~~ {Testing Energy} \\
   
    \midrule
    {WBC}    & {17.00}~~~ {6.200}  & \review{2.917}~~~ \review{0.328} \\
    {MNIST}  & {2162}~~~ {1337}  & \review{118.828}~~~ \review{35.503} \\
    {CARDIO} & {50.70}~~~ {42.30} & \review{9.328}~~~ \review{1.92} \\
    {LYMPHO} & {5.400}~~~ {3.500} & \review{0.536}~~~ \review{0.059} \\
    {SATI2}  & {770.50}~~~ {49.10} & \review{52.878}~~~ \review{1.313} \\
    {MAMMO}  & {1949}~~~ {358.0} & \review{43.663}~~~ \review{0.871} \\

    \bottomrule
    \hline
    \end{tabular}
  \label{tab:energy}
 } 
\end{table*}

We also evaluated the execution time of outlier detection with the different models and datasets. \rev{Table \ref{tab:imodhd_breakdown_training} and Table \ref{tab:imodhd_breakdown_testing} present a breakdown of latency/energy with \immodel~during the different phases of training and testing, respectively. For instance, the \textbf{encoding} of samples into HVs is required for both training and testing phases and requires permutation to be performed on the seed HVs, followed by the bundling operation. While bundling can be performed entirely within the PEs in the \immodel~architecture, permutation requires transfers to/from registers A and B placed at the mat level in our CiM architecture. During training (the most expensive portion of \model), on average across all datasets, encoding accounts for \review{$\sim$88$\%$/$\sim$68.55$\%$ }of the latency/energy. Communication from PEs to registers A and B, and vice-versa, dominate the costs of encoding, accounting for $\sim$58.7$\%$/$\sim$63.2$\%$ of its latency/energy.} 

\rev{Since the latency/energy of encoding during testing is still  (\review{$\sim$99.31$\%$/$\sim$98$\%$}, on average, with respect to the total testing latency/energy), the cost of outlier detection with \immodel~is insignificant, due to the need for successive shifts and additions in the implementation of the pop-count operation in the threshold calculation. Communication from PEs to registers A and B, and vice-versa, during encoding, is similar to the training phase, accounting for $\sim$58.7$\%$/$\sim$63.1$\%$ of the encoding latency/energy.}

Table~\ref{tab:execution_time} shows the execution time for training and testing including the GPU implementation of \model~and the CiM-based implementation (\immodel), along with other baseline methods. In general, conventional models execute faster outlier inference than HDC-based models on the CPU. With a significant amount of cores and faster data transmission between memory and computing unit, GPU achieves lower execution time. However, the HDC-based model still takes a longer time to train and infer than conventional models, e.g., OCSVM and isolation forest.

Our proposed \immodel~ significantly accelerates both the training and testing phases of outlier detection. According to Table \ref{tab:execution_time}, \immodel~shows on average \review{331.5}$\times$ speedup in training and \review{889}$\times$ speedup in inference than \model~running on GPU (the fastest implementation). The training time of \immodel~is slightly large due to the working principle of \immodel~that is amiable with CiM architecture, yet shows extensively superior performance since it is minimal compared to other baseline models for outlier detection. It is challenging to train the MNIST dataset because the model must learn a representation of the input images that is resilient to changes in writing style, stroke thickness, and other elements that can impact how the digits appear. \immodel~can completely learn this dataset in \review{31.67ms}, with an inference time of \review{10.37}ms whereas isolation forest takes 355ms/198ms to train/test on the same dataset. Small datasets like Lympho can be learned in \review{104$\mu$s} and infer any outlier in \review{16.1$\mu$s} using \immodel.

Last, energy results are reported in Table \ref{tab:energy} for the training and testing phases of \immodel. Due to highly parallel calculation in \immodel~fabric, the energy consumption, which factors in both power and latency, is advantageous compared to the GPU-based implementation of \model. On average, energy improvement for \immodel~is at \review{14.0}$\times$/\review{36.9}$\times$ for the training/testing phase.

%The results indicate that the hardware/software codesign implementation of \model~running on our CiM architecture outperforms the GPU-based implementation of the same algorithm by at least 293$\times$/570$\times$ in the training/testing phases in terms of delay.
%So far all the hardware accelerators have been used for machine learning problems like classification, clustering ETC. To make a fair comparison for outlier detection we implement our one class HDC based ODHD in a GPU and compare the energy efficiency achieved by the accelerators. Table~\ref{tab: modified_table} shows the summary for energy consumption for CiM based~\model~and GPU based ODHD. Even though GPUs are highly popular for the parallel computation of the tasks, our CiM architecture with double sensing mechanism makes the algorithm opportune for implementing a highly energy efficient hardware. 

\section{Related Work}
\label{sec:related_work}

In this section, we review related work on models for outlier detection and hardware accelerators for HDC.

\subsection{Models for Outlier Detection}

Outlier detection has been a heavily researched topic with various statistical and machine learning methods proposed. 
One widely-used outlier detection method is the Exemplar-Based Gaussian Mixture Model (GMM) proposed by Yang et al.~\cite{yang2009outlier}, which utilizes a globally optimal expectation maximization (EM) algorithm to fit the GMM to the given dataset. Tang et al.~\cite{tang2015outlier} further improved this method by combining GMM with locality-preserving projections. Another approach uses linear regression, such as the method proposed by Satman et al.~\cite{satman2013new}, which detects outliers based on a non-interactive covariance matrix and concentration steps applied in the least trimmed square estimation.
However, despite their mathematical robustness, statistical methods' assumptions and dependence on a particular distribution model may limit their practical use. \model ~provides a novel approach to outlier detection that does not rely on specific distributional assumptions, making it a promising alternative to existing methods.

Three widely used machine learning-based outlier detection methods are OCSVM, isolation forest, and autoencoder. OCSVM separates outliers from inliers by maximizing the margin and detects samples outside the estimated region as outliers~\cite{li2016anomaly}. In isolation forest, outliers are detected by examining the path length, as they are more sensitive to isolation and have a relatively short traversal path length~\cite{liu2008isolation}. Autoencoder, a neural network-based method, consists of an encoding network and a decoding network. The encoder maps input samples to a low-dimensional feature space, while the decoder reconstructs the sample from the encoded feature. Autoencoder is trained to minimize the reconstruction error and preserve information relevant to normal instances. Outliers, which diverge from the majority of training samples, are hardly reconstructed and lead to a high reconstruction error. Thus, the outliers can be detected by examining the reconstruction error~\cite{he2020exploring}. Despite the popularity of these methods, they rely on different assumptions and may not perform well in various applications.

In recent years, several methods have been proposed for anomaly detection using HDC. One such method, HDAD~\cite{wang2021brief}, adopts an autoencoder-like approach to reconstruct the input samples and detect anomalies based on reconstruction error. However, this method requires tedious encoding and decoding processes, making the detection process cumbersome. In contrast, \model ~proposes a one-class HDC approach for outlier detection, which is fundamentally different from HDAD. We evaluate the performance of \model ~against four baseline methods, namely OCSVM, isolation forest, autoencoder, and HDAD, and provide comprehensive comparison results.

\subsection{CiM Accelerators for HDC}

HDC with its inherent memory-centric operations motivates to implement it in CiM since data movement reduction can be achieved by HV computations fully in memory. \rev{Nevertheless, recent research on DRAM-based CiM designs is tailored to parallel Boolean bitwise operations and often lacks comprehensive support for all operations integral to ODHD. For example, AMBIT~\cite{seshadri2017ambit}, with triple-row activation can execute bitwise majority function but misses native shift operation support essential for HDC encoding. DRISA~\cite{li2017drisa} allows for shift operations within subarrays at the cost of area overhead with multiple microarchitectures for data movement making them inadequate for host memory. DRAM-based CiM architectures, while offering increased computational speed, entail a substantial overhead in terms of processing time, usually requiring several hundred clock cycles for operations involving inputs exceeding three bits. While this approach may be well-suited for tasks like image classification, it may not be a viable choice when designing the architecture for applications involving HVs of 10,000 dimensions. \review{CiM architectures based on lookup tables (LUTs) within DRAM enable fast operations while preserving application level accuracy (e.g.,~\cite{deng2019lacc,sutradhar2020ppim}). However, the use of LUT-based CiM in DRAM may face challenges in managing the size and volume of LUTs required for performing operations in HDC, since the Hypervectors (HV) involved in the computations have thousands of dimensions (10,000+), which requires further investigation.}  }

\rev{The outlined issues with DRAM motivate the search for CiM architectures based on non-volatile memories (NVMs) and CMOS-based SRAMs (our work), which could support more intricate operations.} For the former, Imani et al.~\cite{imani2019searchd} proposed SearcHD, which utilizes the analog properties of ReRAM-based in-memory computing (IMC) arrays to employ HD blocks in memory with a fully binarized computing algorithm. However, the energy and time required to program the MAJ IMC array from the XOR IMC array severely limit their ability to be used efficiently. In our work, we distribute each bipolar HV of D dimensions across the readily available technology CMOS-based SRAM in a holistic way reducing data transfer overheads. Leveraging from the digital domain computation without any ADC/DAC or current controlled PEs, computation is fully exerted in memory using elements with smaller hardware footprints. By realizing training and testing phases without using analog operations, our CiM architecture improves the time complexity and energy consumption without trading off reliability, which makes it a good fit for low-power hardware devices, aligned with other proposed architectures~\cite{eggimann20215}.

\section{Conclusion}
\label{sec:conclusion}
In this study, we propose \model, a novel outlier detection algorithm based on hyperdimensional computing (HDC), a non-traditional machine learning paradigm. Additionally, we present \immodel, a computing-in-memory (CiM) hardware and software (HW/SW) co-design implementation to enhance latency and energy efficiency. The proposed \model ~algorithm leverages a learning structure to generate a one-class hypervector (HV) based on inlier samples. This HV represents the abstract information of all inlier samples, and any testing sample with an HV dissimilar from this HV is identified as an outlier. Both the training and testing phases of \model ~can be performed using conventional CPU/GPU hardware or our proposed SRAM-based CiM architecture using HW/SW co-design techniques. We evaluate the performance of \model~on six datasets from different application domains using three metrics -- accuracy, F1 score, and ROC-AUC and compare it with several baseline methods, such as OCSVM, isolation forest, and autoencoder. The experimental results show that \model ~outperforms all the baseline methods in terms of these three metrics on every dataset for both CPU/GPU and CiM implementations. Moreover, we conduct an extensive design space exploration to demonstrate the tradeoff between delay, energy efficiency, and performance of \model. We show that \immodel, the in-memory computing-based implementation of \model, outperforms the GPU-based implementation of \model~ by at least \review{331.5$\times$/889}$\times$ in terms of training/testing latency and on average \review{14.0$\times$/36.9$\times$} in terms of training/testing energy consumption.

\section*{Acknowledgement}

\bibliography{ODHD}

\begin{thebibliography}{10}
\providecommand{\url}[1]{#1}
\csname url@samestyle\endcsname
\providecommand{\newblock}{\relax}
\providecommand{\bibinfo}[2]{#2}
\providecommand{\BIBentrySTDinterwordspacing}{\spaceskip=0pt\relax}
\providecommand{\BIBentryALTinterwordstretchfactor}{4}
\providecommand{\BIBentryALTinterwordspacing}{\spaceskip=\fontdimen2\font plus
\BIBentryALTinterwordstretchfactor\fontdimen3\font minus
  \fontdimen4\font\relax}
\providecommand{\BIBforeignlanguage}[2]{{%
\expandafter\ifx\csname l@#1\endcsname\relax
\typeout{** WARNING: IEEEtran.bst: No hyphenation pattern has been}%
\typeout{** loaded for the language `#1'. Using the pattern for}%
\typeout{** the default language instead.}%
\else
\language=\csname l@#1\endcsname
\fi
#2}}
\providecommand{\BIBdecl}{\relax}
\BIBdecl

\bibitem{tang2015outlier}
X.-m. Tang, R.-x. Yuan, and J.~Chen, ``Outlier detection in energy
  disaggregation using subspace learning and gaussian mixture model,''
  \emph{International Journal of Control and Automation}, vol.~8, no.~8, pp.
  161--170, 2015.

\bibitem{latecki2007outlier}
L.~J. Latecki, A.~Lazarevic, and D.~Pokrajac, ``Outlier detection with kernel
  density functions,'' in \emph{MLDM}, vol.~7, 2007, pp. 61--75.

\bibitem{li2016anomaly}
Y.~Li, T.~Zhang, Y.~Y. Ma, and C.~Zhou, ``Anomaly detection of user behavior
  for database security audit based on ocsvm,'' in \emph{2016 3rd International
  Conference on Information Science and Control Engineering (ICISCE)}.\hskip
  1em plus 0.5em minus 0.4em\relax IEEE, 2016, pp. 214--219.

\bibitem{liu2008isolation}
F.~T. Liu, K.~M. Ting, and Z.-H. Zhou, ``Isolation forest,'' in \emph{2008
  eighth ieee international conference on data mining}.\hskip 1em plus 0.5em
  minus 0.4em\relax IEEE, 2008, pp. 413--422.

\bibitem{he2020exploring}
T.~He, L.~Zhang, F.~Kong, and A.~Salekin, ``Exploring inherent sensor
  redundancy for automotive anomaly detection,'' in \emph{2020 57th ACM/IEEE
  Design Automation Conference (DAC)}.\hskip 1em plus 0.5em minus 0.4em\relax
  IEEE, 2020, pp. 1--6.

\bibitem{kanerva2009hyperdimensional}
P.~Kanerva, ``Hyperdimensional computing: An introduction to computing in
  distributed representation with high-dimensional random vectors,''
  \emph{Cognitive computation}, vol.~1, pp. 139--159, 2009.

\bibitem{ge2020classification}
L.~Ge and K.~K. Parhi, ``Classification using hyperdimensional computing: A
  review,'' \emph{IEEE Circuits and Systems Magazine}, vol.~20, no.~2, pp.
  30--47, 2020.

\bibitem{hersche2020integrating}
M.~Hersche, E.~M. Rella, A.~Di~Mauro, L.~Benini, and A.~Rahimi, ``Integrating
  event-based dynamic vision sensors with sparse hyperdimensional computing: a
  low-power accelerator with online learning capability,'' in \emph{Proceedings
  of the ACM/IEEE International Symposium on Low Power Electronics and Design},
  2020, pp. 169--174.

\bibitem{elkan2008learning}
C.~Elkan and K.~Noto, ``Learning classifiers from only positive and unlabeled
  data,'' in \emph{Proceedings of the 14th ACM SIGKDD international conference
  on Knowledge discovery and data mining}, 2008, pp. 213--220.

\bibitem{ielmini2020device}
D.~Ielmini and G.~Pedretti, ``Device and circuit architectures for in-memory
  computing,'' \emph{Advanced Intelligent Systems}, vol.~2, no.~7, p. 2000040,
  2020.

\bibitem{wang2022odhd}
R.~Wang, X.~Jiao, and X.~S. Hu, ``{ODHD}: one-class brain-inspired
  hyperdimensional computing for outlier detection,'' in \emph{Proceedings of
  the 59th ACM/IEEE Design Automation Conference}, 2022, pp. 43--48.

\bibitem{Rayana2016}
\BIBentryALTinterwordspacing
S.~Rayana, ``Outlier detection datasets (odds) library,'' 2016. [Online].
  Available: \url{http://odds.cs.stonybrook.edu}
\BIBentrySTDinterwordspacing

\bibitem{kang14}
M.~Kang, M.~S. Keel, N.~R. Shanbhag, S.~Eilert, and K.~Curewitz, ``{An
  energy-efficient VLSI architecture for pattern recognition via deep embedding
  of computation in SRAM},'' in \emph{The International Conference on
  Acoustics, Speech, and Signal Processing}, 2014.

\bibitem{yin2018ultra}
X.~Yin, K.~Ni, D.~Reis, S.~Datta, M.~Niemier, and X.~S. Hu, ``An ultra-dense
  2fefet tcam design based on a multi-domain fefet model,'' \emph{IEEE
  Transactions on Circuits and Systems II: Express Briefs}, vol.~66, no.~9, pp.
  1577--1581, 2018.

\bibitem{kazemi2022achieving}
A.~Kazemi, F.~M{\"u}ller, M.~M. Sharifi, H.~Errahmouni, G.~Gerlach,
  T.~K{\"a}mpfe, M.~Imani, X.~S. Hu, and M.~Niemier, ``Achieving
  software-equivalent accuracy for hyperdimensional computing with
  ferroelectric-based in-memory computing,'' \emph{Scientific reports},
  vol.~12, no.~1, p. 19201, 2022.

\bibitem{liu19_hdc_rram}
J.~Liu, M.~Ma, Z.~Zhu, Y.~Wang, and H.~Yang, ``{HDC-IM}: Hyperdimensional
  computing in-memory architecture based on rram,'' in \emph{2019 26th IEEE
  International Conference on Electronics, Circuits and Systems (ICECS)}, 2019,
  pp. 450--453.

\bibitem{reis2018computing}
D.~Reis, M.~Niemier, and X.~S. Hu, ``Computing in memory with fefets,'' in
  \emph{Proceedings of the international symposium on low power electronics and
  design}, 2018, pp. 1--6.

\bibitem{reis20_date}
D.~Reis, A.~F. Laguna, M.~Niemier, and X.~S. Hu, ``A fast and energy efficient
  computing-in-memory architecture for few-shot learning applications,'' in
  \emph{2020 Design, Automation \& Test in Europe Conference \& Exhibition
  (DATE)}, 2020, pp. 127--132.

\bibitem{aga17}
S.~{Aga}, S.~{Jeloka}, A.~{Subramaniyan}, S.~{Narayanasamy}, D.~{Blaauw}, and
  R.~{Das}, ``Compute caches,'' in \emph{2017 IEEE International Symposium on
  High Performance Computer Architecture (HPCA)}, Feb 2017.

\bibitem{reis20_tvlsi}
D.~Reis, J.~Takeshita, T.~Jung, M.~Niemier, and X.~S. Hu, ``Computing-in-memory
  for performance and energy-efficient homomorphic encryption,'' \emph{IEEE
  Transactions on Very Large Scale Integration (VLSI) Systems}, vol.~28,
  no.~11, pp. 2300--2313, 2020.

\bibitem{kim2018efficient}
Y.~Kim, M.~Imani, and T.~S. Rosing, ``Efficient human activity recognition
  using hyperdimensional computing,'' in \emph{Proceedings of the 8th
  International Conference on the Internet of Things}, 2018, pp. 1--6.

\bibitem{wang2021brief}
R.~Wang, F.~Kong, H.~Sudler, and X.~Jiao, ``Brief industry paper: {HDAD}:
  Hyperdimensional computing-based anomaly detection for automotive sensor
  attacks,'' in \emph{2021 IEEE 27th Real-Time and Embedded Technology and
  Applications Symposium (RTAS)}.\hskip 1em plus 0.5em minus 0.4em\relax IEEE,
  2021, pp. 461--464.

\bibitem{fujiki2019duality}
D.~Fujiki, S.~Mahlke, and R.~Das, ``Duality cache for data parallel
  acceleration,'' in \emph{Proceedings of the 46th International Symposium on
  Computer Architecture}, 2019, pp. 397--410.

\bibitem{Ranjan2019XMANNAC}
A.~Ranjan, S.~Jain, J.~R. Stevens, D.~Das, B.~Kaul, and A.~Raghunathan,
  ``{X-MANN}: A crossbar based architecture for memory augmented neural
  networks,'' \emph{2019 56th ACM/IEEE Design Automation Conference (DAC)}, pp.
  1--6, 2019.

\bibitem{zimek2013subsampling}
A.~Zimek, M.~Gaudet, R.~J. Campello, and J.~Sander, ``Subsampling for efficient
  and effective unsupervised outlier detection ensembles,'' in
  \emph{Proceedings of the 19th ACM SIGKDD international conference on
  Knowledge discovery and data mining}, 2013, pp. 428--436.

\bibitem{sathe2016lodes}
S.~Sathe and C.~Aggarwal, ``{LODES}: Local density meets spectral outlier
  detection,'' in \emph{Proceedings of the 2016 SIAM international conference
  on data mining}.\hskip 1em plus 0.5em minus 0.4em\relax SIAM, 2016, pp.
  171--179.

\bibitem{wang2018hyperparameter}
S.~Wang, Q.~Liu, E.~Zhu, F.~Porikli, and J.~Yin, ``Hyperparameter selection of
  one-class support vector machine by self-adaptive data shifting,''
  \emph{Pattern Recognition}, vol.~74, pp. 198--211, 2018.

\bibitem{hwinfo}
``{HWiNFO - Free System Information Monitoring and Diagnostics},'' [online]
  Available: \url{https://www.hwinfo.com/}.

\bibitem{mo2023haac}
J.~Mo, J.~Gopinath, and B.~Reagen, ``Haac: A hardware-software co-design to
  accelerate garbled circuits,'' in \emph{Proceedings of the 50th Annual
  International Symposium on Computer Architecture}, 2023, pp. 1--13.

\bibitem{maxwell2021using}
P.~Maxwell, D.~Niblick, and D.~C. Ruiz, ``Using side channel information and
  artificial intelligence for malware detection,'' in \emph{2021 IEEE
  International Conference on Artificial Intelligence and Computer Applications
  (ICAICA)}.\hskip 1em plus 0.5em minus 0.4em\relax IEEE, 2021, pp. 408--413.

\bibitem{heddes2022torchhd}
M.~Heddes, I.~Nunes, P.~Verg{\'e}s, D.~Desai, T.~Givargis, and A.~Nicolau,
  ``Torchhd: An open-source python library to support hyperdimensional
  computing research,'' \emph{arXiv preprint arXiv:2205.09208}, 2022.

\bibitem{wang2020further}
Z.~Wang, B.~Dai, D.~Wipf, and J.~Zhu, ``Further analysis of outlier detection
  with deep generative models,'' 2020.

\bibitem{everingham2010pascal}
M.~Everingham, L.~Van~Gool, C.~K. Williams, J.~Winn, and A.~Zisserman, ``The
  pascal visual object classes (voc) challenge,'' \emph{International journal
  of computer vision}, vol.~88, pp. 303--338, 2010.

\bibitem{poremba2015destiny}
M.~Poremba, S.~Mittal, D.~Li, J.~S. Vetter, and Y.~Xie, ``Destiny: A tool for
  modeling emerging 3d nvm and edram caches,'' in \emph{2015 Design, Automation
  \& Test in Europe Conference \& Exhibition (DATE)}.\hskip 1em plus 0.5em
  minus 0.4em\relax IEEE, 2015, pp. 1543--1546.

\bibitem{zhao2007predictive}
W.~Zhao and Y.~Cao, ``Predictive technology model for nano-cmos design
  exploration,'' \emph{ACM Journal on Emerging Technologies in Computing
  Systems (JETC)}, vol.~3, no.~1, pp. 1--es, 2007.

\bibitem{knudsen2008nangate}
J.~Knudsen, ``Nangate 45nm open cell library,'' \emph{CDNLive, EMEA}, 2008.

\bibitem{burrello2018one}
A.~Burrello, K.~Schindler, L.~Benini, and A.~Rahimi, ``One-shot learning for
  ieeg seizure detection using end-to-end binary operations: Local binary
  patterns with hyperdimensional computing,'' in \emph{2018 IEEE Biomedical
  Circuits and Systems Conference (BioCAS)}.\hskip 1em plus 0.5em minus
  0.4em\relax IEEE, 2018, pp. 1--4.

\bibitem{yang2009outlier}
X.~Yang, L.~J. Latecki, and D.~Pokrajac, ``Outlier detection with globally
  optimal exemplar-based gmm,'' in \emph{Proceedings of the 2009 SIAM
  international conference on data mining}.\hskip 1em plus 0.5em minus
  0.4em\relax SIAM, 2009, pp. 145--154.

\bibitem{satman2013new}
M.~H. Satman, ``A new algorithm for detecting outliers in linear regression,''
  \emph{International Journal of statistics and Probability}, vol.~2, no.~3, p.
  101, 2013.

\bibitem{seshadri2017ambit}
V.~Seshadri, D.~Lee, T.~Mullins, H.~Hassan, A.~Boroumand, J.~Kim, M.~A. Kozuch,
  O.~Mutlu, P.~B. Gibbons, and T.~C. Mowry, ``Ambit: In-memory accelerator for
  bulk bitwise operations using commodity dram technology,'' in
  \emph{Proceedings of the 50th Annual IEEE/ACM International Symposium on
  Microarchitecture}, 2017, pp. 273--287.

\bibitem{li2017drisa}
S.~Li, D.~Niu, K.~T. Malladi, H.~Zheng, B.~Brennan, and Y.~Xie, ``Drisa: A
  dram-based reconfigurable in-situ accelerator,'' in \emph{Proceedings of the
  50th Annual IEEE/ACM International Symposium on Microarchitecture}, 2017, pp.
  288--301.

\bibitem{deng2019lacc}
Q.~Deng, Y.~Zhang, M.~Zhang, and J.~Yang, ``{LAcc}: Exploiting lookup
  table-based fast and accurate vector multiplication in dram-based cnn
  accelerator,'' in \emph{Proceedings of the 56th Annual Design Automation
  Conference 2019}, 2019, pp. 1--6.

\bibitem{sutradhar2020ppim}
P.~R. Sutradhar, M.~Connolly, S.~Bavikadi, S.~M.~P. Dinakarrao, M.~A. Indovina,
  and A.~Ganguly, ``{pPIM}: A programmable processor-in-memory architecture
  with precision-scaling for deep learning,'' \emph{IEEE Computer Architecture
  Letters}, vol.~19, no.~2, pp. 118--121, 2020.

\bibitem{imani2019searchd}
M.~Imani, X.~Yin, J.~Messerly, S.~Gupta, M.~Niemier, X.~S. Hu, and T.~Rosing,
  ``Searc{HD}: A memory-centric hyperdimensional computing with stochastic
  training,'' \emph{IEEE Transactions on Computer-Aided Design of Integrated
  Circuits and Systems}, vol.~39, no.~10, pp. 2422--2433, 2019.

\bibitem{eggimann20215}
M.~Eggimann, A.~Rahimi, and L.~Benini, ``A 5 $\mu$w standard cell memory-based
  configurable hyperdimensional computing accelerator for always-on smart
  sensing,'' \emph{IEEE Transactions on Circuits and Systems I: Regular
  Papers}, vol.~68, no.~10, pp. 4116--4128, 2021.

\end{thebibliography}

%https://tex.stackexchange.com/questions/374612/reduce-the-gap-between-bios-in-ieeetran

\vskip -2\baselineskip plus -1fil
\begin{IEEEbiography}[{\includegraphics[width=1in,height=1.25in,clip,angle=0]{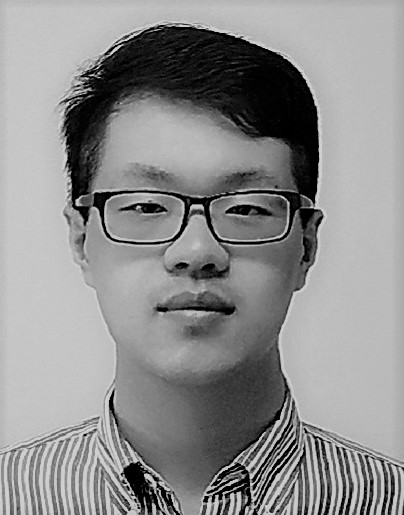}}]%
{Ruixuan Wang} is a Ph.D. candidate in Computer Engineering (CpE) from the Department of Electrical and Computer Engineering at Villanova University. In 2020 he received his M.Sc. degree in Computer Engineering at New York University, U.S. His research interest includes deep learning, approximate computing, hyperdimensional computing, machine learning security and robustness.

\end{IEEEbiography}
%Use this to reduce the huge amounts of space between bios
\vskip -2\baselineskip plus -1fil
\begin{IEEEbiography}[{\includegraphics[width=1in,height=1.25in,clip]{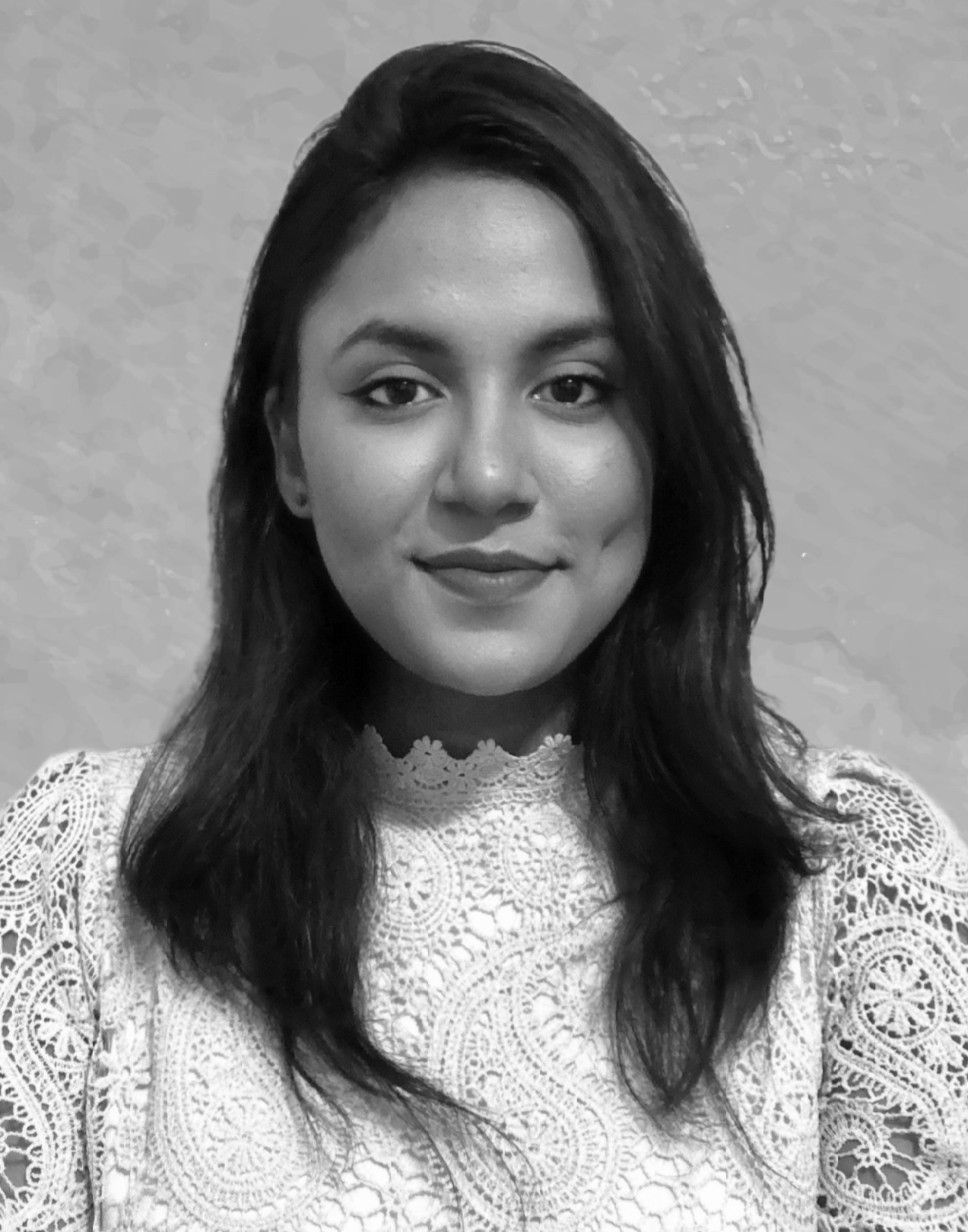}}]%
{Sabrina Hassan Moon}
is currently pursuing her Ph.D. in Computer Science and Engineering at the University of South Florida. Her research focuses on computing in memory, hardware-software co-design for machine learning applications, emerging devices, device characterization, and VLSI. She received her B.S. degree from Shahjalal University of Science and Technology in Bangladesh. Sabrina is a devoted individual committed to promoting women`s contributions in academia.
\end{IEEEbiography}
%Use this to reduce the huge amounts of space between bios
\vskip -2\baselineskip plus -1fil
\begin{IEEEbiography}[{\includegraphics[width=1in,height=1.25in,clip]{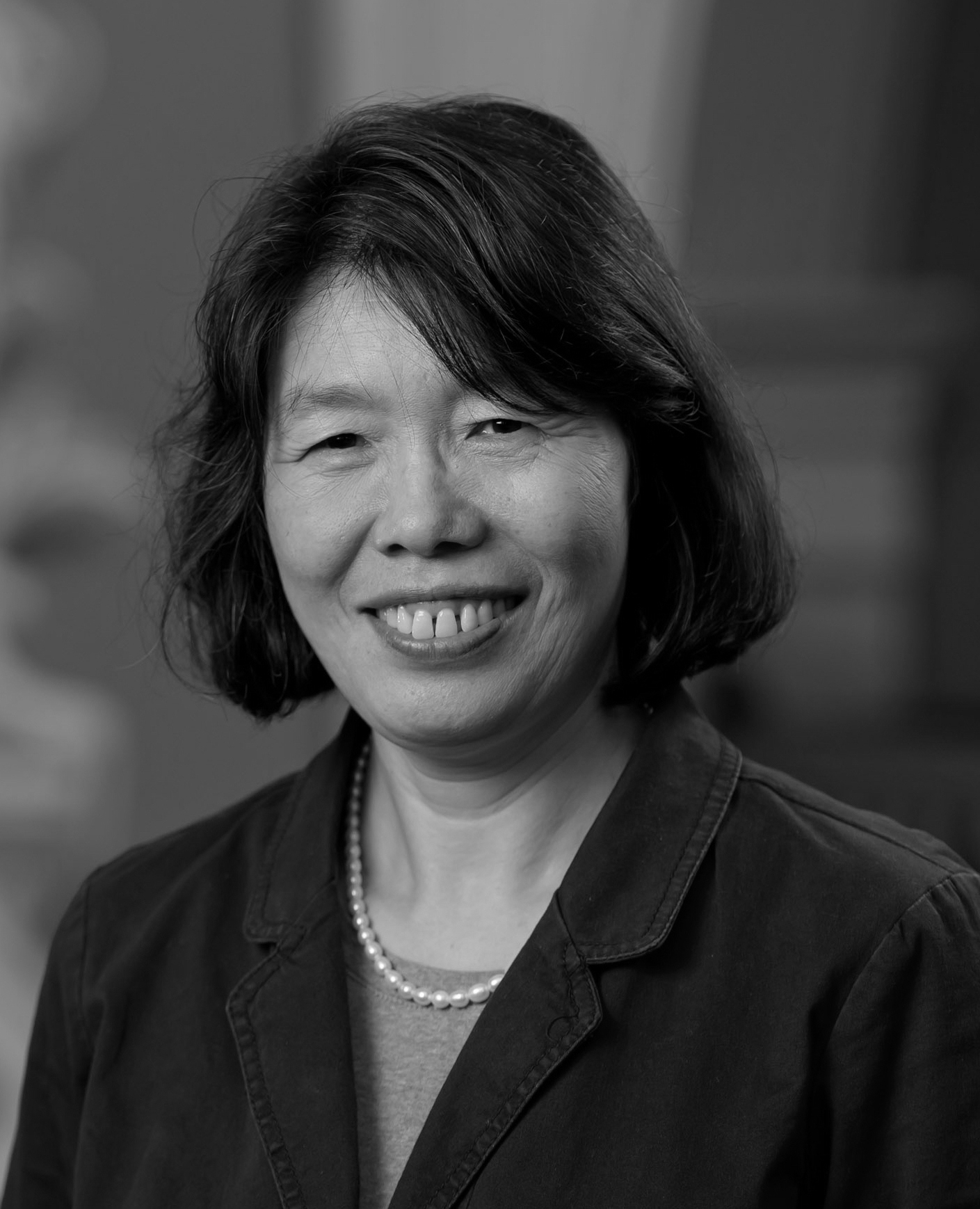}}]%
{Xiaobo Sharon Hu}
(S'85-M'89-SM'02-F'16) received her Ph.D. from Purdue University, M.S. from Polytechnic Institute of New York, and B.S. from Tianjin University. She is a Professor at the University of Notre Dame. Her research interests include energy/reliability-aware system design, circuit and architecture design with emerging technologies, real-time embedded systems, and hardware-software co-design. 
%She is the Editor-in-Chief of ACM Transactions on Design Automation of Electronic Systems, and also served as Associate Editor for IEEE Transactions on CAD, IEEE TVLSI,  ACM Transactions on Embedded Computing, etc. 
She received the NSF CAREER Award in 1997, the Best Paper Award from Design Automation Conference in 2001, ACM/IEEE International Symposium on Low Power Electronics and Design in 2018, etc. 
\end{IEEEbiography}
%Use this to reduce the huge amounts of space between bios
\vskip -2\baselineskip plus -1fil
\begin{IEEEbiography}[{\includegraphics[width=1in,height=1.25in,clip]{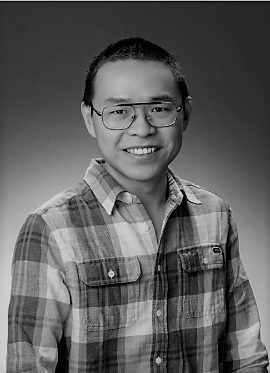}}]%
{Xun Jiao} is an assistant professor in the ECE Department at Villanova University. He was a visiting scientist of Meta/Facebook, and a visiting student researcher of NXP Semiconductors. He received his Ph.D. from UC San Diego in 2018, and B.S. degree from Beijing University of Posts and Telecommunications in 2013. His research interests include software-hardware codesign, design automation, bio-inspired computing, and machine learning, with a particular focus on designing robust and energy-efficient systems. %He published more than 50 papers in first-tier journals and conferences. He is an Associate Editor of IEEE Trans on CAD and a program committee member of conferences such as DAC, ICCAD, and ASP-DAC. He has received 6 best paper awards/nominations in conferences such as DATE, EMSOFT, and SELSE. He is the recipient of the IEEE Young Electrical Engineer of the Year Award (Philadelphia Section). He has delivered an invited presentation in U.S. Congressional House. 
His research is funded by NSF, NIH, and industry corporations (L3Harris, NVIDIA).

\end{IEEEbiography}
\vskip -2\baselineskip plus -1fil
\begin{IEEEbiography}[{\includegraphics[width=1in,height=1.25in,clip]{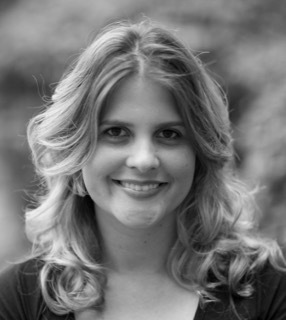}}]%
{Dayane Reis}
is an Assistant Professor in the Department of CSE at the University of South Florida. She received her Ph.D. from the University of Notre Dame, M.S. from the Federal University of Minas Gerais, Brazil, and B.S. from PUC-MG, Brazil. Her research focuses on the design of circuits and architectures for data-intensive computing. Dayane was one of the best paper award winners in the ACM/IEEE International Symposium on Low Power Electronics and Design, and a recipient of the Cadence Women in Technology (WIT) Scholarship 2018/2019.
\end{IEEEbiography}

\end{document}